\documentclass[preprint,3p]{elsarticle}      

\usepackage{amsmath}
\usepackage{natbib}
\usepackage{amssymb}
\usepackage{graphicx}
\usepackage{dcolumn}
\usepackage{bm}
\usepackage{mathrsfs}
\usepackage{color}
\usepackage{psfrag}
\usepackage{url}
\usepackage{soul}
\usepackage{float}  
\usepackage[normalem]{ulem}
\usepackage[algo2e, ruled, lined, vlined, longend]{algorithm2e}

\usepackage{caption}
\usepackage{subcaption}
\usepackage[normalem]{ulem}
\usepackage{color}

\definecolor{revColor}{rgb}{0.12,0.45,0.0}
\definecolor{revdColor}{rgb}{0.8,0.15,0.25}

\usepackage{booktabs}
  
\usepackage[pdftitle={The PDF title},
  pdffitwindow=true,
  breaklinks=true,
  colorlinks=true,
  urlcolor=blue,
  linkcolor=black,
  citecolor=black,
  anchorcolor=red]{hyperref}
  
\usepackage{graphicx}
\usepackage{subcaption}

\usepackage{tikz}
\usetikzlibrary{shapes.geometric, arrows}

\tikzstyle{rect} = [draw, rectangle, text width=2.5cm, minimum height=1cm, text centered, draw=black, fill=white!20]
\tikzstyle{rect2} = [rectangle, text width=2.3cm, minimum height=1cm, text centered, draw=black, fill=white!20]
\tikzstyle{line} = [draw, -latex']

\begin{document}

\title{In-situ Estimation of Time-averaging Uncertainties in Turbulent Flow Simulations}

\author[focal1,focal2,focal3]{S. Rezaeiravesh\corref{cor1}}
\ead{saleh.rezaeiravesh@manchester.ac.uk}
\author[focal4]{C. Gscheidle\corref{cor2}}
\ead{christian.gscheidle@scai.fraunhofer.de}
\author[focal2,focal3]{A. Peplinski}
\ead{adam@mech.kth.se}
\author[focal4,focal5]{J. Garcke}
\ead{jochen.garcke@scai.fraunhofer.de}
\author[focal6,focal2,focal3]{P. Schlatter}
\ead{philipp.schlatter@fau.de}
\cortext[cor1]{Principal Corresponding Author}
\cortext[cor2]{Corresponding Author}
\address[focal1]{Department of Fluids and Environment, The University of Manchester, M139PL Manchester, UK}
\address[focal2]{SimEx/FLOW, Engineering Mechanics, KTH Royal Institute of Technology, 10044 Stockholm, Sweden}
\address[focal3]{Swedish e-Science Research Centre (SeRC), Stockholm, Sweden}
\address[focal4]{Fraunhofer SCAI, 53757 Sankt Augustin, Germany} 
\address[focal5]{Institute for Numerical Simulation, University of Bonn, 53115 Bonn, Germany} 
\address[focal6]{Institute of Fluid Mechanics (LSTM), Friedrich--Alexander Universit\"at Erlangen--N\"urnberg (FAU), 91058 Erlangen, Germany}

\date{\today}


\newcommand{\et}{et al.}
\newcommand{\eg}{e.g.}
\newcommand{\ie}{i.e.}
\newcommand{\fig}{Figure}
\newcommand{\figs}{Figures}
\newcommand{\eq}{Eq.}
\newcommand{\eqs}{Eqs.}
\newcommand{\sect}{Section}
\newcommand{\alg}{Algorithm}
\newcommand{\sects}{Sections}
\newcommand{\rf}{Ref.}
\newcommand{\rfs}{Refs.}

\newcommand{\fx}{\mathbf{x}}
\newcommand{\bx}{\bar{x}}

\newcommand{\hmu}{\hat{\mu}}
\newcommand{\hsig}{\hat{\sigma}}
\newcommand{\BE}{\mathbb{E}}
\newcommand{\hBE}{\hat{\mathbb{E}}}
\newcommand{\BV}{\mathbb{V}}
\newcommand{\hBV}{\hat{\mathbb{V}}}
\newcommand{\cN}{\mathcal{N}}
\newcommand{\cU}{\mathcal{U}}
\newcommand{\cov}{\rm{cov}}
\newcommand{\hcov}{\widehat{\rm{cov}}}
\newcommand{\train}{{\rm {train}}}

\newcommand{\rey}{{\rm Re}}
\newcommand{\reyt}{{\rm Re}_\tau}

\begin{abstract}
The statistics obtained from turbulent flow simulations are generally uncertain due to finite time averaging. 
The techniques available in the literature to accurately estimate these uncertainties typically only work in an offline mode, that is, they require access to all available samples of a time series at once. 
In addition to the impossibility of online monitoring of  uncertainties during the course of simulations, such an offline approach can lead to input/output (I/O) deficiencies and large storage/memory requirements, which can be problematic for large-scale simulations of turbulent flows. 
Here, we designed, implemented and tested a framework for estimating time-averaging uncertainties in turbulence statistics in an in-situ (online/streaming/updating) manner. 
The proposed algorithm relies on a novel low-memory update formula for computing the sample-estimated autocorrelation functions (ACFs). Based on this, smooth modeled ACFs of turbulence quantities can be generated to accurately estimate the time-averaging uncertainties in the corresponding sample mean estimators. 
The resulting uncertainty estimates are highly robust, accurate, and quantitatively the same as those obtained by standard offline estimators.
Moreover, the computational overhead added by the in-situ algorithm is found to be negligible. 
The framework is completely general and can be used with any flow solver and also integrated into the simulations over conformal and complex meshes created by adopting adaptive mesh refinement techniques.
The results of the study are encouraging for the further development of the in-situ framework for other uncertainty quantification and data-driven analyses relevant not only to large-scale turbulent flow simulations, but also to the simulation of other dynamical systems leading to time-varying quantities with autocorrelated samples.
\end{abstract}
\maketitle

\section{Introduction}
Turbulent fluid flows are fundamentally unsteady and contain vortical structures of a wide range of spatial and temporal scales. 
For numerical simulation of turbulent flows, various approaches have been developed, see \eg~\citet{sagaut:13}. 
Due to their capabilities in accurately capturing the physics of turbulent flows, scale-resolving approaches such as large-eddy simulation (LES) and direct numerical simulation (DNS) are of particular interest for both academic and industrial flows. 
It is recalled that LES, mostly, and DNS, fully, resolve the flow structures in time and space. 
A main challenge when applying these approaches is the required excessive computational cost, which would become prohibitive for wall-bounded turbulent flows at high-Reynolds numbers~\cite{deck:14}. 
For such flows, which appear in many engineering applications, the computational cost of the scale-resolving simulations is driven by the requirement of accurately resolving the inner part of the turbulent boundary layers (TBL)~\cite{choi:12,jimenez:13}.
In recent decades, the progress in high-performance computing (HPC) technologies has made it possible to employ scale-resolving turbulence simulations such as LES and DNS at higher Reynolds numbers and more complex flows.

The HPC developments have, however, led to new requirements and aspects to consider in the design of the next generation of CFD (computational fluid dynamics) software. 
In particular, the NASA 2030 vision~\cite{nasa:2030} has listed a technology development roadmap and specified the readiness level of various technologies. 
Among such are the uncertainty quantification (UQ) techniques for assessing the reliability and accuracy of the CFD simulations' outcomes.
Among others, the uncertainties in CFD simulations can be due to the numerical settings, programming, turbulence modeling, initial/boundary data, and finite time-averaging~\cite{lesReli1,smith:13,oliver:14,uqFrame:22}. 
The focus of the present study is on the latter that is also known as statistical or sampling uncertainty, appearing due to the finite number of samples considered when computing the time-averaged quantities and turbulence statistics. 
In practice, after the turbulent flow becomes statistically stationary, sample mean estimators (SME) are evaluated by averaging the samples which are autocorrelated by nature. 
Different techniques for estimating the uncertainty in SMEs of turbulent flow quantities have been used, see \cite{oliver:14,russo:17,xavier:21,lenschow:94}, a short review of which is given in \sect~\ref{sec:uqOverview}. 
As extensively discussed in \rf~\cite{xavier:21}, the hyperparameters appearing in each of these techniques have to be properly adopted in order to avoid any bias in the estimated time-averaging uncertainties.

In large-scale CFD simulations, there is an increasing gap between the amount of data that is generated during the runtime and what can actually be stored to disk. For current HPC systems, this gap due to the limitations in the data input/output (I/O) can be as large as up to four orders of magnitudes, even for highly parallel systems~\cite{khan:21}. To overcome this limitation, workflows for in-situ visualizations have been set up to export compressed pictures of the flow field during runtime, see~\cite{atzori:22} for an in-situ visualization workflow with Nek5000~\cite{Fischer2008}. A common technique to visualize turbulence are iso-surfaces of the so-called Q-criterion exported for each time-step, which can later be integrated into a video. Even though the resolution and the number of extracted pictures can be increased significantly by an in-situ visualization workflow, a quantitative analysis based on these will not be possible after the end of the simulation. 
Moreover, in the field of data analytics and machine learning, in-situ algorithms, also known as updating, streaming or incremental, have recently gained attention for two reasons. First, there are cases where not all of the training data fits into memory. Therefore, the data processing is usually performed out-of-memory, where only one sample or a small batch of samples is loaded and processed at a time to update the model. This also applies for scenarios where data is constantly being produced, \eg~by a simulation or measurements of a system.
But here, as soon as new data is available it can be processed step by step in an in-situ fashion. 
Note that when processing data in batches, often the accuracy of the results depends on the batch size, \eg~when computing streaming singular-value decomposition (SVD)~\cite{liang:16}. In these cases, a good balance between memory usage on the HPC system and the accuracy of the results is required.
Second, when working in-situ one can exploit the intermediate data analysis results or measurements for an online monitoring of the system. 
Both motivations also apply to CFD simulations. There, very large amounts of data are generated that cannot be held in memory at once. 
An in-situ analysis can thus significantly increase the accuracy and accessibility to the data analysis results. 
Besides, based on an analysis of intermediate simulation results, we can build a monitoring system that provides criteria to abort or steer the simulation and eventually save computing time and resources. 

In the literature, studies for the evaluation of time-averaging uncertainties have essentially been based on the assumption that the uncertainty estimators have access to the whole set of time samples, see \cite{oliver:14,russo:17,xavier:21}.
In practice, this requires that for each quantity, the generated samples are stored on disk and then loaded to the memory.
These would lead to both severe memory and computational deficiencies, especially for large-scale simulations. To rectify these issues, the present study provides the necessary tools in addition to a framework for an in-situ evaluation of the  uncertainties in sample mean estimators for autocorrelated samples. 
These include updating estimators for the SME and associated uncertainties, as well as an interface between the CFD solver and UQ module. 
The memory requirements will be minimized in approaches where the UQ estimator still needs to keep information in memory at runtime.

The remainder of the paper is organized as follows. 
\sect~\ref{sec:uqOverview} introduces the problem of estimating time-averaging uncertainty in SMEs of turbulence time series and proposes a formula for obtaining a smooth model for the autocorrelation function (ACF) of such time series.
The focus of \sect~\ref{sec:upUQMethods} is on updating UQ algorithms.
\sect~\ref{sec:upMS} reviews the updating formulae for an in-situ estimation of the sample mean and variance of a time series. 
The formulae are then used in \sect~\ref{sec:batchMethods} to derive expressions for an in-situ estimation of time-averaging uncertainty in SMEs based on batch-based methods. 
The updating method for estimating the time-averaging uncertainties proposed in the present study is discussed in \sect~\ref{sec:iMACF}, followed by the description of the workflow and software that implements it in \sect~\ref{sec:workflow}.
Then, in \sect~\ref{sec:CFD}, the flow solvers and use cases to which the framework is applied are described.
In \sect~\ref{sec:results}, the results are presented and thoroughly discussed. 
This includes the assessment of the accuracy, sensitivity and robustness of the new techniques proposed in \sects~\ref{sec:uqOverview} and ~\ref{sec:iMACF}, as well as the assessment of the computational efficiency of the framework developed in \sect~\ref{sec:workflow} when applied to the use cases. 
Summary and conclusions are provided in \sect~\ref{sec:summary}.

\section{Uncertainty in a Sample Mean Estimator (SME)}\label{sec:uqOverview}
In practice when only one simulation (realization) of a physical process is performed, the true expectation or mean value of a quantity~$x$ belonging to the process cannot be obtained.
Instead, for each realization of the process, a set of time samples for~$x$ is obtained from which a sample mean estimate can be computed.
Let~$\fx$ denote a time series sample of size~$n$, \ie~$\fx=\{x_i\}_{i=1}^n$, where $x_i=x(t_i)$. Henceforth, we assume the times~$t_i$ are equi-spaced. 
The sample mean estimator (SME) for the time series reads as
\begin{equation}\label{eq:sme}
    \hmu:=\hBE[x]=\frac{1}{n}\sum_{i=1}^n x_i \,.
\end{equation}
Note that throughout the text, estimated values/estimators are represented with the overhat symbol~$\hat{\cdot}$. 
The SME is unbiased, see \eg~\cite{wei:06}, and converges to the true but unobserved expectation of~$x$, \ie~$\mu:=\BE[x]$. 
Furthermore, the SME has a Gaussian distribution:
\begin{equation}
    \hmu\sim \cN\left(\mu,\sigma^2(\hmu) \right)\,,
\end{equation}
where~$\sigma(\hmu)$ is the standard deviation and hence a measure of uncertainty of~$\hmu$.
For many processes including flow turbulence, the time samples of any flow variable are autocorrelated over a generally unknown number of time lags. 
For autocorrelated samples in~$\fx$, an analytical expression can be derived for the variance of~$\hmu$, see \eg~\cite{wei:06}:
\begin{equation}\label{eq:varMu}
    \BV[\hmu]=\sigma^2(\hmu) =  \frac{1}{n}\left[  \gamma_0+2\sum_{m=1}^{(n-1)} \left(1-\frac{m}{n}\right) \gamma_m\right] \,,
\end{equation}
where~$\gamma_m$ is the autocovariance of~$x$ at lag~$m$. 
Trivially, this expression can be written in terms of the autocorrelations $\rho_m=\gamma_m/\gamma_0$ with~$\gamma_0$ denoting the variance of~$x$.
The sample-estimated autocovariances are obtained from 
\begin{equation}\label{eq:sampleACov}
    \hat{\gamma}_m=\hcov(x_i,x_{i-m}) = \frac{1}{n-m} \sum_{i=1}^{n-m} x'_i x'_{i-m} \,,
\end{equation}
where $x'_i=x_i-\hBE[x]$. 
Plugging~$\hat{\gamma}_m$ into \eq~(\ref{eq:varMu}) leads to estimates for~$\hat{\sigma}^2(\hmu)$, which can, however, be inaccurate noting there are non-vanishing oscillations of~$\hat{\gamma}_m$ at higher lags for any finite number of time samples (sample size).

In the literature, two main approaches have been suggested to circumvent the issues arising by the use of~$\hat{\gamma}_m$ in \eq~(\ref{eq:varMu}). 
In the batch-based approaches, the original samples are divided into a set of batches and then one works with the mean values of the samples in each batch without using \eq~(\ref{eq:varMu}).
Depending on how particularly the batch means are used, different approximations of~$\hat{\sigma}(\hmu)$ are achieved. 
For a review of the batch-based uncertainty estimators as well as the more efficient Batch-Means Batch-Correlation (BMBC) algorithm, see \cite{russo:17}.
In \sect~\ref{sec:batchMethods}, we provide a brief overview of the batch-based methods relevant to the present study.

In the second type of approaches, \eq~(\ref{eq:varMu}) is directly evaluated by introducing a modeled autocorrelation function (ACF) or autocovariance function (ACovF) that is smooth for all lags.
As first applied to turbulent flow simulations by \citet{oliver:14}, an autoregressive model (ARM) can be fitted to the original samples~$\fx$, and then the estimated ARM coefficients are used to fit a smooth power-law ACF for~$m\in[0,\infty)$, see \ref{app:ARM_modeledACF}.
This approach is believed to be the most accurate method of estimating uncertainty in the SME in turbulent flows conditioned on adopting appropriate values for the hyperparameters involved. 
The latter include the order of the ARM and the set of sample-estimated autocorrelations to construct a smooth ACF model, see \citet{xavier:21}.
The described method is not of interest in the present study, because it requires the entire time-series data to be in memory for fitting the ARM and is therefore unsuitable for an in-situ implementation. 
Instead, we can rely on ad-hoc algebraic functions to create smooth models for the ACFs, meaning that fitting an ARM to the time samples is skipped. 
Along this line of thought, in the studies mainly relevant to atmospheric turbulence the following function proposed in \rf~\cite{lenschow:94} is widely used to model the ACFs:
\begin{equation}\label{eq:MACF_fun1}
    \rho(m) = \exp(-a\,m) \,,
\end{equation}
for $m=0,1,2,\ldots$, where~$1/a$ is the turbulence integral time-scale. 
There have been other forms of functions for modeling ACF, see \eg~\cite{salesky:12} and the references therein. 
However, as discussed in \sect~\ref{sec:accuracyMACF}, the model function (\ref{eq:MACF_fun1}) may not be appropriate for data obtained from wall-bounded turbulence simulations. 
Instead, we propose to use the slightly modified function 
\begin{equation}\label{eq:acfModel}
    \rho(m) = a\exp(-b\, m) + (1-a) \exp(-c\, m) \,,
\end{equation}
to model ACFs of turbulence time series.
Here,~$m\geq 0$ is the time lag (integer), and parameters~$b$ and~$c$ are strictly positive. 
Obviously,~$\rho(0)=1$ and~$\rho (m\to\infty)\to 0$.
In comparison to \eq~(\ref{eq:MACF_fun1}), the suggested function has two more parameters to estimate. 
To construct a smooth ACF model using \eq~(\ref{eq:acfModel}), a set of training sample-estimated autocorrelations $\{\rho_i\}_{i=1}^{n_\train}$ at lags $\{m_i\}_{i=1}^{n_\train}$ are used. 
As discussed in \sect~\ref{sec:accuracyMACF}, as one of its main advantages, the model function~(\ref{eq:acfModel}) is flexible in terms of how the training lags are selected. 
To estimate the model parameters~$a$,~$b$, and~$c$, a non-linear least-squares method can be employed. 
Hereafter, we refer to the estimation of~the~$\sigma(\hmu)$ using the ACF model~(\ref{eq:acfModel}) as the \emph{MACF method}.

\section{Updating UQ Algorithms}
\label{sec:upUQMethods}
To the author's knowledge, there has not been any prior study addressing the in-situ formulation of neither batch-based methods nor estimators relying on the modeling of the ACF function. 
In what follows, first the basic definitions for constructing updating uncertainty estimators are provided. 
Then, updating versions of the batch-based methods as well as the proposed MACF method are presented. 
In all the following derivations, the samples are assumed to be equidistant in time. 
For the special case of statistically stationary turbulent flows, the variable~$x$ in the following formulations can be any of the flow variables at any spatial location inside the flow domain.

\subsection{Updating sample mean and variance estimators}\label{sec:upMS}
Following \citet{chan:83}, the sample mean of a time series considering the $i$-th to the $j$-th time samples is defined as
\begin{equation}\label{eq:Mij}
    M_{i,j} = \frac{1}{(j-i+1)} \sum_{k=i}^j x_k \,.
\end{equation}
Correspondingly, the sample variance estimation is given by $S_{i,j}/(j-i+1)$ where, 
\begin{equation}\label{eq:Sij}
    S_{i,j} = \sum_{k=i}^j x_k^2 - (j-i+1) M_{i,j}^2 \,.
\end{equation}
Welford~\cite{welford:62} proposed updating formulations for $M_{i,j}$ and $S_{i,j}$:
\begin{eqnarray}
  M_{i,j} &=& M_{i,j-1} + \frac{1}{(j-i+1)} \left(x_j-M_{i,j-1} \right) = M_{i,j-1} + \Delta M_{i,j}\,, \label{eq:upM}\\
  S_{i,j} &=& S_{i,j-1} + (j-i)(j-i+1) \Delta M^2_{i,j} \,.  \label{eq:upS}
\end{eqnarray}
These formulae are the building blocks for the updating batch-based estimators of~$\sigma(\hmu)$, as detailed below.

\subsection{Updating batch-based methods}\label{sec:batchMethods}
\begin{figure}
    \centering
    \includegraphics[scale=0.6]{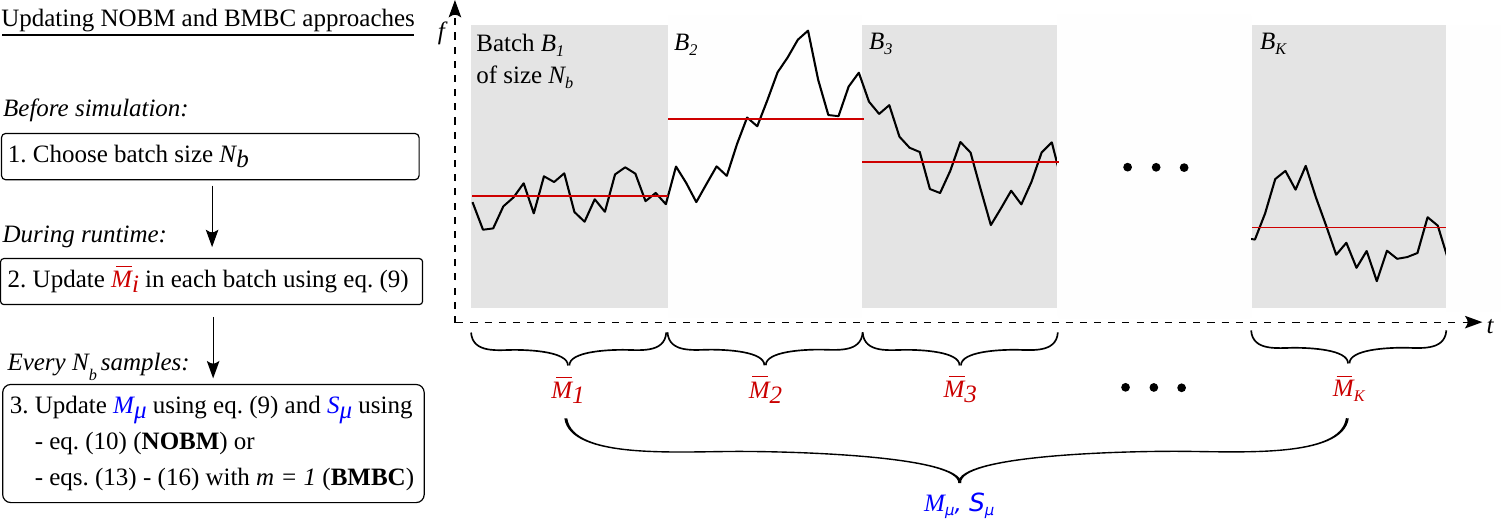}
    \caption{Schematic representation of the updating algorithm applied to the NOBM and BMBC methods. The batch means and the SME of~$x$  are updated using \eq~(\ref{eq:upM}). The variance of the SME for the NOBM and BMBC methods is computed using \eqs~(\ref{eq:upS}) and~(\ref{eq:bmbcVar}), respectively.}
    \label{fig:schem_nobm}
\end{figure}

In the standard (offline) batch-based methods, see~\rf~\cite{russo:17}, first a set of batches are created using the time samples collected from a turbulence time series. 
For the non-overlapping batch mean (NOBM) method that we discuss here, the set of samples $\{x_i\}_{i=1}^n$ are divided into~$K$ batches of size~$N_b$. 
Then,~$\bx_k$ for $k=1,2,\ldots,K$ is computed as the sample mean of the samples included in the~$k$-th batch.
Using $\{\bx_k\}_{k=1}^K$, the sample mean of the original time series is estimated by
\begin{equation}
    \hmu=\frac{1}{K}\sum_{k=1}^K \bx_k \,,
\end{equation}
the variance of which is estimated as
\begin{equation}\label{eq:nobmVar}
\hsig^2(\hmu)=\frac{1}{K(K-1)}\sum_{k=1}^K (\bx_k-\hmu)^2 \,.
\end{equation}
Performing these steps in an in-situ (updating) manner is straightforward. 
The main point is that two sets of sample means should be updated using \eq~(\ref{eq:upM}): one for computing the batch-means, and the other one for computing the sample mean of the time series using the updated batch-means. 
But \eq~(\ref{eq:upS}) is used only ``per request" to update the estimated variance of the SME of the time series. 
These steps are summarized below and also schematically represented in \fig~\ref{fig:schem_nobm}.

\begin{enumerate}
    \item Specify the batch size $N_b$; 
    \item In the flow simulation's time loop, with a given sampling interval:
    \begin{enumerate}
       \item compute sample mean of every~$N_b$ samples using updating formula~(\ref{eq:upM}) $\rightarrow \bar{M}_i$; 
       \item update the SME of the time series and its variance via \eqs~(\ref{eq:upM}) and (\ref{eq:upS}), respectively $\rightarrow M_\mu ,S_\mu$.
    \end{enumerate}
\end{enumerate}

To improve the accuracy of~$\hsig(\hmu)$ estimated by the NOBM method, Russo and Luchini~\cite{russo:17} suggested considering the first-lag correlation between the batch means. 
The resulting batch-means and batch-correlation (BMBC) estimator for the SME uncertainty reads as, 
\begin{equation}\label{eq:bmbcVar}
\hsig^2(\hmu)= \frac{1}{(K-1)(K-2)} 
\left(
\sum_{k=1}^K (\bx^2_k - \hmu^2)
+2 \sum_{k=1}^{K-1} (\bx_k \bx_{k+1} - \hmu^2)
\right) \,,
\end{equation}
which, compared to~\eq~(\ref{eq:nobmVar}), has the second summation on its right-hand-side (RHS). 
In order to formulate an updating formula for this term, \eq~(\ref{eq:upACF}) below can be used with~$m=1$ and substituting~$x$ by~$\bar{x}$. 
The implementation of the updating BMBC (iBMBC) method requires the last computed batch mean to be stored in memory.

\subsection{Updating MACF (iMACF) method}
\label{sec:iMACF}
As a main drawback, the accuracy of the NOBM and BMBC methods is controlled by the batch size~$N_b$, which, in general, cannot be chosen intuitively prior to the simulations. 
Moreover, the optimal value of~$N_b$ depends on the flow variable and also changes with the spatial location~\cite{xavier:21}.
These restrict the suitability of adopting the batch-based uncertainty estimators for in-situ applications.
Therefore, we focus on the approach of modeling the autocorrelation~$\rho_m=\gamma_m/\gamma_0$ in~\eq~(\ref{eq:varMu}) for any lag~$m>0$.
In particular, the proposed MACF method outlined in \sect~\ref{sec:uqOverview} is considered, where a set of modeling parameters are estimated using the training sample-estimated ACF values~$\{\rho_i\}_{i=1}^{n_\train}$ corresponding to the time lags~$\{m_i\}_{i=1}^{n_\train}$. 
The only potential subjectivity in this method could be due to the choice of the training data set which, however, as discussed in \sect~\ref{sec:accuracyMACF}, does only have a negligible influence on the accuracy of the predicted uncertainties. This is a clear indication of the robustness of the proposed ACF model~(\ref{eq:acfModel}).

The standard estimation of the sample autocovariance at lag~$m\geq 0$ via \eq~(\ref{eq:sampleACov}) using~$n$ time samples requires having all~$n-m$ samples at once. 
Instead, we aim at deriving a formulation for sample-estimated ACF which is updating and requires a minimum amount of storage. 
As the first step, we define the sample-estimated ACovF at lag~$m$ by including the samples from~$i$ to~$j$, as:
\begin{equation}\label{eq:acov_op}
\tilde{\Gamma}^m_{i,j} = \Gamma^m_{i,j}/(j-i+1)     \,,
\end{equation}
where,
\begin{equation}\label{eq:Gamma_ij}
\Gamma^m_{i,j} = \sum_{k=i}^j (x_k-M_{i,j})(x_{k-m}-M_{i,j}) \,,
\end{equation}
where $M_{i,j}$ is the updating SME of~$x$ defined in \eq~(\ref{eq:Mij}).
Expanding this definition results in the following
expression:
\begin{eqnarray}\label{eq:upACF}
    \Gamma^m_{i,j} &=&
    \Gamma^m_{i,j-1} 
    - \Delta M_{i,j} \sum_{k=i}^{j-1}(x_k+x_{k-m}) 
    + (j-i-m+1) M_{i,j}^2 
    - (j-i-m) M_{i,j-1}^2 \nonumber \\
    && +~ x_j x_{j-m} 
    - M_{i,j} (x_j+x_{j-m}) \,,
\end{eqnarray}
where~$i$ is, in practice, fixed (\ie~chosen after the statistically stationary condition of the turbulent flow is established), and for a given~$m$ it is needed that~$i>m$. 
As detailed in \ref{app:acf2var}, for $m=0$ this expression becomes identical to \eq~(\ref{eq:upS}) for the variance of~$x$.
For computing all the terms on the RHS of \eq~(\ref{eq:upACF}) except the last two, no storage of the samples of~$x$ is needed. 
The storage demanded by the two terms containing~$x_{j-m}$ can basically increase with~$j$ and~$m$, however, our optimal implementation requires a buffer of a fixed size. 
In fact, this size can be kept small even for obtaining a high accuracy of the estimated uncertainty of an SME, thanks to the flexibility of the uncertainty estimation techniques described below.
In principle, going from $j-1$ to~$j$, first the sample at~$j-m$ that is already available in the buffer is used to update the last two terms in \eq~(\ref{eq:upACF}).
Then, the oldest stored sample is removed from the buffer followed by adding the most recent sample~$x_j$ to the buffer. 

Our investigation in \sect~\ref{sec:accuracyMACF} shows that, for accurately modeling the ACF of a time series using \eq~(\ref{eq:acfModel}), only a few sample-estimated autocorrelations are required. 
Moreover, there is a great flexibility regarding the selection of the training samples. 
Therefore, \eq~(\ref{eq:upACF}) should be evaluated at a set of $m \in \mathbf{m}_\train$. 
Bearing in mind the issue with~$x_{j-m}$, the maximum value of~$m$, hereafter~$m_{\max}$, can be a main factor in driving the overall cost of the in-situ estimation of the sample-estimated autocorrelations using \eq~(\ref{eq:upACF}).

In practice, to enhance the computational efficiency and minimize the required memory storage of the iMACF method for in-situ estimation of the uncertainty of an SME, see \alg~\ref{alg:iMACF}, time samples can be taken with a frequency of~$T_s \Delta t$, where,~$T_s$ is a positive integer and~$\Delta t$ is the time-step size in the flow simulation. 
The selection $\mathbf{m}_\train$ can be linked to~$T_s$ to create an efficient computational algorithm.

    
    

\begin{algorithm2e}[!tb]
  \DontPrintSemicolon
  \SetAlgoLined
  \SetNoFillComment
  \LinesNumbered
  \SetArgSty{textnormal}

\caption{Updating MACF (iMACF) method.}
 \label{alg:iMACF}

   \KwIn{Choose~$m_{\max}$ and~$T_s$, where~$m_{\max}$ is sufficiently larger than~$T_s$. Although these choices are, in general, arbitrary, choose the~$m_{\max}$ such that $m_{\max}=M_{\max} T_s$ with $M_{\max}$ being an integer.
    This makes the initialization step more efficient.}

  \vspace{0.3cm}

 Initialize the list of training lags as $\mathbf{m}_{\train} = [T_s,2T_s,\ldots,M_{\max} T_s]$. 
    For each $m\in \mathbf{m}_{\train}$, a buffer array is created to store the corresponding~$\Gamma^m_{i,j}$.\;
    
  \While{CFD simulation is not finished}{
    \If{time~$t$ is divisible by~$T_s$ }{
        \uIf{$t\leq m_{\max}=M_{\max}T_s$}{
            fill the buffer array with the samples of~$x$.
        }
        \Else{
            Compute $\Delta M_{i,j}$ in \eq~(\ref{eq:upM}),\;
            Update $S_{i,j}$ using \eq~(\ref{eq:upS}),\;
            \ForAll{$m \in \mathbf{m}_{\train}$}{
                update $\Gamma_{ij}^m$ using \eq~(\ref{eq:upACF}),
            }
            Update the buffer array,\;
            Update $M_{i,j}$ in \eq~(\ref{eq:upM}).\;
            (optional) Use the training autocorrelations at lags $ \mathbf{m}_{\train}$ to fit the ACF model (\ref{eq:acfModel}) which is then plugged into \eq~(\ref{eq:varMu}) to estimate $\hsig(\hmu)$.\;
        }
    }
  }
\end{algorithm2e}

\begin{figure}[t]
    \centering
    \includegraphics[scale=0.75]{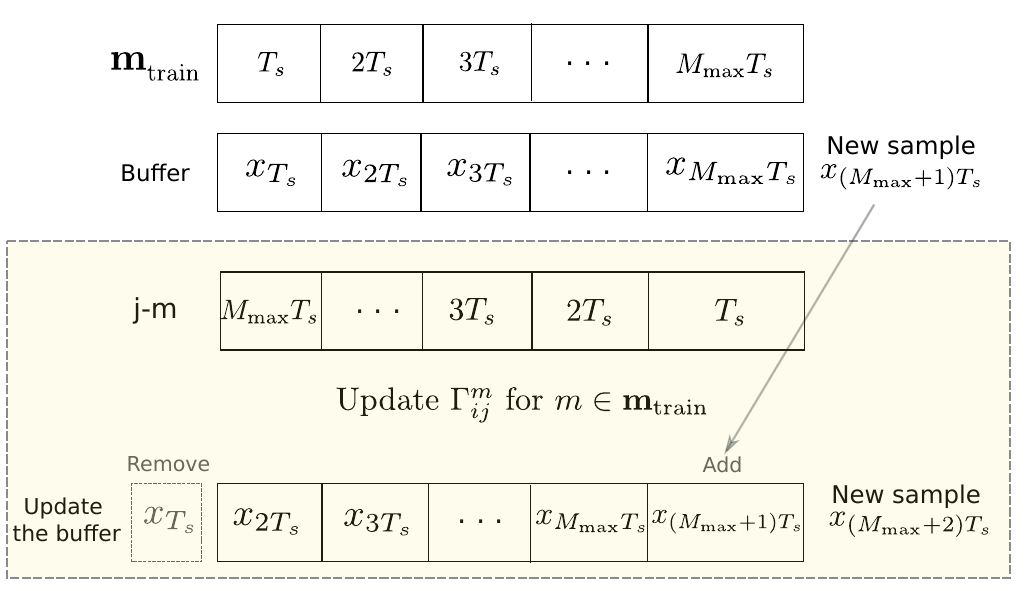}
    \caption{Schematic representation of the updating algorithm to compute sample autocovariances from \eq~(\ref{eq:acov_op}) using the updating expression~(\ref{eq:upACF}), see \alg~\ref{alg:iMACF}. The operations within the shaded area are repeated in a loop over $m\in\mathbf{m}_{\rm train}$.}
    \label{fig:iMACF}
\end{figure}

Line 14 in \alg~\ref{alg:iMACF} is executed upon a user request or automatically following a preset schedule, as this step is used for monitoring the convergence of the turbulence statistics. 
The procedure for computing the sample training autocovariances is shown schematically in \fig~\ref{fig:iMACF}. 
Recall that the buffer is required for the purpose of in-situ evaluation of the terms containing~$x_{j-m}$ in \eq~(\ref{eq:upACF}). 
The low-storage feature of the algorithm can be inferred from Line 12: the sample at the smallest lag in the array is removed, other samples are shifted, and the new sample is appended to the end of the buffer array. 
However, to increase the efficiency in practice, the oldest value in the buffer is overwritten by the newest value while keeping all other values untouched, resulting in a cyclic writing to the buffer array. 
In particular, for the Python library Numpy~\cite{numpy} used in the present work, this has been implemented by using numpy.roll which creates copies of the buffer array for each execution. 

\section{Workflow Design and Software Implementation}\label{sec:workflow}
The updating UQ methods from \sect~\ref{sec:upUQMethods} are implemented in Python as an extension of UQit~\cite{uqit} to allow for a flexible adaptation of the algorithms during the development and validation of the framework. 
In this section, the proposed workflow applied to the CFD simulations and its components are described. 

The CFD simulation mesh is processed block-wise on each MPI (Message Passing Interface) rank to take advantage of the matrix manipulations in linear algebra libraries, more specifically Numpy~\cite{numpy}, and provide a straight-forward high-level parallelization of our code. 
For the purpose of benchmarking the algorithms in a realistic simulation environment, we utilize the ParaView Catalyst~\cite{catalyst} interface. 
Therefore, algorithms are wrapped as VTK (Visualization Toolkit) filters and executed in a Catalyst pipeline together with native VTK filters, for instance, to extract slices from the original mesh. Although this setup is generally independent of the flow solver, the tests in the following sections were performed using Nek5000~\cite{Fischer2008} with an additional Catalyst interface~\cite{catalyst}. 
The performance analysis of the entire in-situ workflow is discussed in \sect~\ref{sec:inSituPerf}. 

Referring to \sect~\ref{sec:iMACF}, we note that the algorithm can be split into three major parts: the updating formula to compute the ACF at different lags, the estimation of the continuous ACF via curve-fitting (\ie~\eq~(\ref{eq:acfModel})), and finally the estimation of uncertainty in the SME of turbulence quantities. 
While all three steps could, in principle, be executed during the runtime of a simulation, we may prefer to run the second and third steps offline. 
This allows us to take advantage of the reduced data I/O due to the online computation of sample-estimated ACF with only a small amount of additional runtime and memory allocation. 
On the other hand, the UQ results are only needed at the end of the simulation or at a few intermediate time steps preset or requested by a user. 
Thus, the computationally intense curve-fitting step can be detached from the workflow to avoid slowing down the simulation for which the in-situ UQ is enabled. 
Following this, we implemented the second and third steps of the algorithm in a ParaView plugin to further process the data that is generated during the simulation. 
Note that if the computational architecture allows it, idle computational resources can be used for the detached steps, see~\cite{ju:2022}. 

\begin{figure}[t]

    \centering
    \includegraphics[scale=0.57]{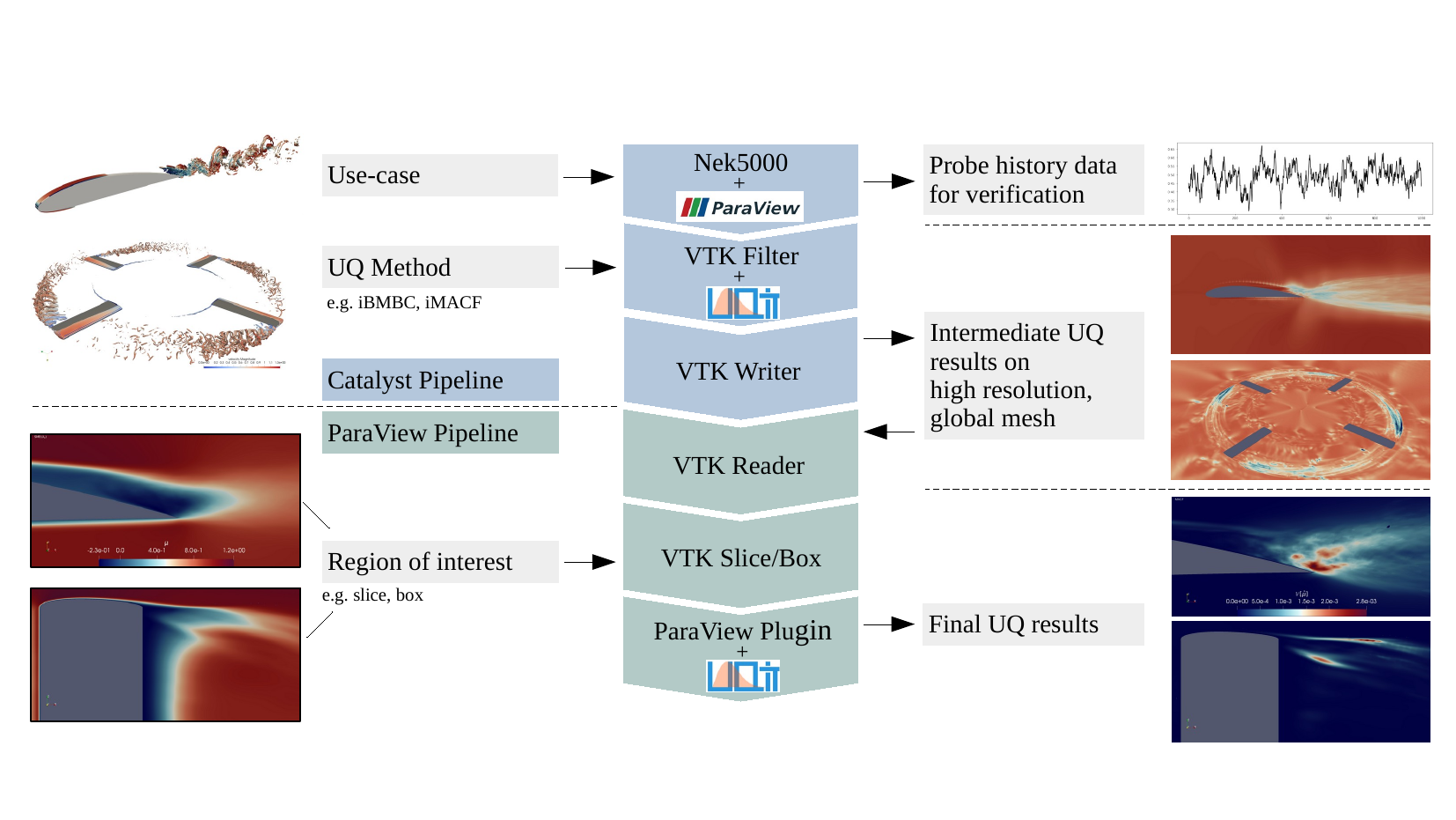}    
    \caption{The workflow of the in-situ UQ framework proposed in \sect~\ref{sec:iMACF} to be applied to CFD simulations.}\label{fig:workflow}
\end{figure}

The verification of the framework has been carried out by comparing the proposed updating algorithm to the corresponding offline algorithms from the statsmodels Python library~\cite{statsmodels}, which has access to the full time series, using the data from a DNS of the 3D flow around a cylinder at $\rey=3900$ as introduced in~\sect~\ref{sec:CFD}.
For this purpose, we set up a Catalyst workflow to extract a 2D slice from the original mesh that gets both processed in-situ by the described algorithm and written to disk in a VTK format, at the same time steps. That way, we can process the exact same data in both online and offline modes, see \fig~\ref{fig:workflow}.
For test data from the cylinder use case with a set of~$100000$ time-steps, both results are identical up to around~$14$ significant figures, providing a strong indication that the implementation of the in-situ UQ algorithm is practically error-free. 
As there are many additions of values that differ by several orders of magnitude during the evaluation of the updating algorithm, we also investigated the influence of reducing the floating point precision to 32-bit. 
In this case, after~$100000$ time-steps the accuracy drops down to around~$4$ significant figures. 
Thus, single precision floats should provide sufficient accuracy of the results for most cases, as the numerical error is still much smaller than the sampling uncertainty in the sample mean estimations.

\section{Flow Solver and Use Cases}\label{sec:CFD}
\subsection{Flow solver}
The described workflow can be applied to any flow solver, however, we consider Nek5000~\cite{Fischer2008} in the present study. 
Nek5000 is an open-source spectral-element~\cite{patera1984spectral} flow solver developed at Argonne National Laboratory (US). 
The low numerical dissipation and high parallel performance~\cite{Offermans2016strong} makes this software perfect for high-fidelity simulation of large-scale advection-diffusion problems, and in particular turbulent flows, see \eg~\rfs~\cite{elkhoury:2013,merzari2020,vinuesa2018turbulent}. 
In the spectral-element method, the computational domain is decomposed into a set of non-overlapping spectral subdomains called elements. 
The Navier--Stokes equations in their weak form are then discretized over such elements treated as spectral domains. We focus here on the staggered grid formulation, so-called $P_N-P_{N-2}$ formulation, in which velocity and pressure are represented on different meshes. 
In this case, the functional spaces of the primitive variables velocity and pressure are spanned by the Lagrangian interpolants on the Gauss–Lobatto–Legendre (GLL) and Gauss–Legendre (GL) quadrature points, respectively. 
A high spatial resolution is obtained by adopting high-order polynomials with order~$N$. 
In the simulations discussed in the present work, we have used~$N$ equal~5 for the rotating parts simulation,~$7$ for the channel flow and~$11$ for the NACA4412 flow.
For integration in time, a semi-implicit scheme is adopted in which the nonlinear terms in the Navier--Stokes equations are treated explicitly and the remaining unsteady Stokes problem is solved implicitly. More in-depth discussion of the algorithm can be found in \eg~\rf~\cite{Fischer2002}.

The domain decomposition into elements is a main source of both the algorithm parallelism and meshing flexibility. 
The latter can be significantly improved by including an adaptive mesh refinement (AMR) strategy, which is a self-adapting algorithm allowing to dynamically modify the mesh according to the estimated computational error. 
An AMR version of Nek5000~\cite{Offermans2020,Offermans2023} was developed at KTH using h-type refinement, in which the total number of grid points is modified by splitting/merging the elements while keeping the number of collocation points per elements fixed. This implementation follows a conforming-space/nonconforming-mesh approach~\cite{Fischer2002}, in which a special interpolation operator is applied at nonconforming interfaces avoiding the construction of the so-called mortar elements~\cite{Maday1989}. 
Although this approach limits the permitted mesh configurations, it allows to use simple and efficient mesh management tools based on the octree refinement (\eg~p4est library~\cite{Burstedde2011}), and has a relatively small impact on the solver parallel performance.
Regarding the AMR simulations in \sect~\ref{sec:AMRrotor}, the mesh refinement was driven by a spectral error indicator formulated by Mavriplis~\cite{Mavriplis1990}.

\subsection{Use cases}\label{sec:useCases}
In this section, the flow cases to which the in-situ UQ framework is applied are briefly introduced.
Due to their complexity and engineering relevance, wall-bounded turbulent flows are considered. 
In order to show the generality of the UQ framework, the simulations include both conformal and non-conformal (AMR) computational grids.

\subsubsection{Simulations with conformal mesh}\label{sec:useConformal}
As one of the most canonical wall-bounded turbulent flows, the turbulent channel flow is considered which comprises of two parallel flat walls separated by distance~$2\delta$.
This internal flow is periodic in the streamwise and spanwise directions, which correspond to~$x$ and~$z$, respectively. 
The wall-normal direction is represented by~$y$.
The absence of uncertainty due to initial and boundary conditions makes the turbulent channel flow suitable for fundamental studies. 
For a-priori assessment of the proposed UQ algorithm, the data of turbulent channel flow at the friction Reynolds number $\rey_\tau=300$ are used in \sects~\ref{sec:accuracyMACF} and ~\ref{sec:validMACF}. 
This Reynolds number is defined as $\reyt=u_\tau \delta/\nu$, where $u_\tau$ is the averaged wall friction velocity and~$\nu$ denotes the kinematic viscosity.
The data includes the wall-normal profile of the streamwise velocity component and wall friction velocity averaged over time and the periodic directions. 
The simulation was performed using polynomial order~$N=7$ in Nek5000 and a total number of~$10^5$ time samples at time intervals~$0.008 \delta/U_b$ ($U_b$ is the bulk velocity) were collected to analyze the UQ algorithm in an offline mode.

\begin{figure}
    \centering
    \begin{tabular}{cc}
       \includegraphics[scale=0.2]{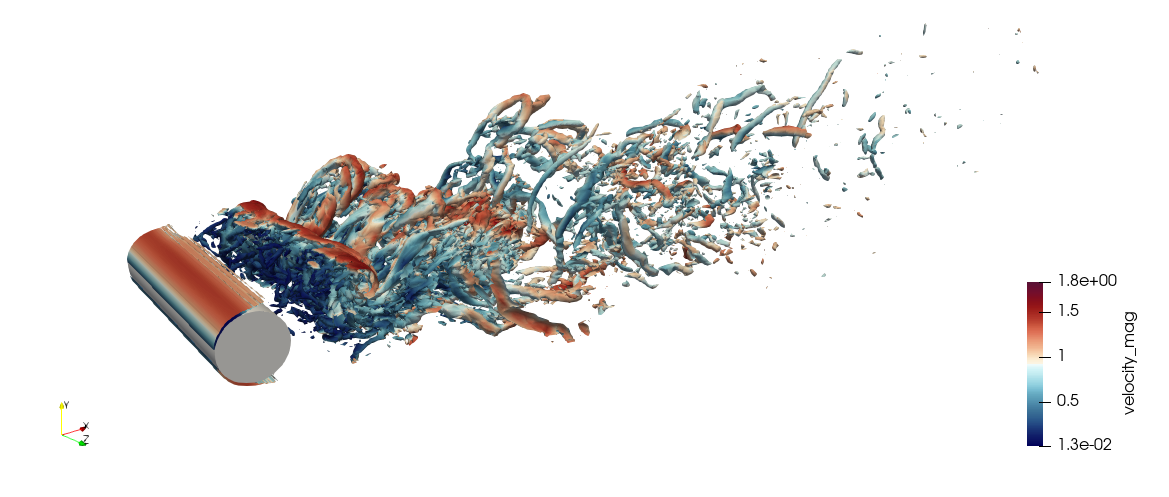} & \hspace{-0.9cm}
       \includegraphics[scale=0.2]{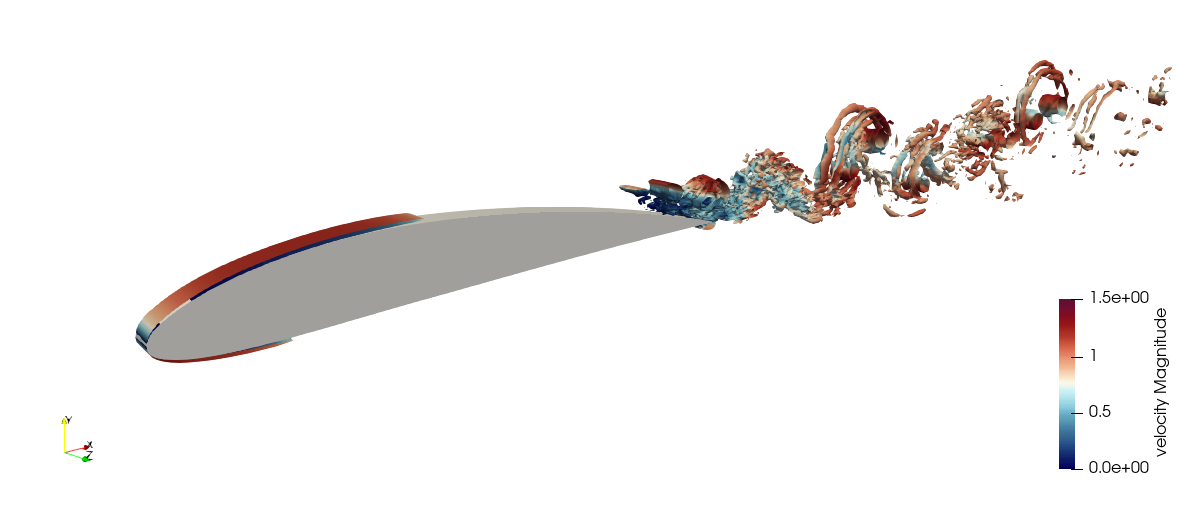}   \\
       {\small (a)} & {\small (b)}\\
    \end{tabular}       
    \caption{Flow around a (a) circular cylinder at $\rey=3900$ and (b) a NACA4412 wing section at $\rey=75000$ computed with Nek5000. Visualization of the instantaneous flow fields using isosurfaces of the Q-criterion colored by the velocity magnitude.}
    \label{fig:cylinder_flow}
\end{figure}

The turbulent flow around a cylinder at $\rey=3900$ as well as a NACA4412 wing-section at $\rey=75000$ is considered as an external aerodynamic use case to which the UQ framework is integrated in an in-situ way.
\fig~\ref{fig:cylinder_flow} represents the iso-surfaces of the Q-criterion of a snapshot of the turbulent flow fields.
In the spanwise direction, a periodic boundary condition is applied. 
The flows contain several interesting flow features, such as laminar separation and periodic structures in the wake region. 
Estimation of the uncertainty in the flow statistics in the wake region is especially challenging due to the strong autocorrelation of the time samples over large time averaging intervals. 
In \sect~\ref{sec:resultsInSitu}, the in-situ UQ framework is applied to estimate the uncertainty in the sample mean velocity at all GLL points of the three-dimensional mesh. 
Furthermore, an analysis of the computational performance of the in-situ UQ framework integrated into the CFD simulations is detailed in \sect~\ref{sec:inSituPerf}.

\subsubsection{AMR simulation of a rotor}\label{sec:AMRrotor}
\begin{figure}
    \centering
    \begin{tabular}{cc}
       \includegraphics[scale=0.09]{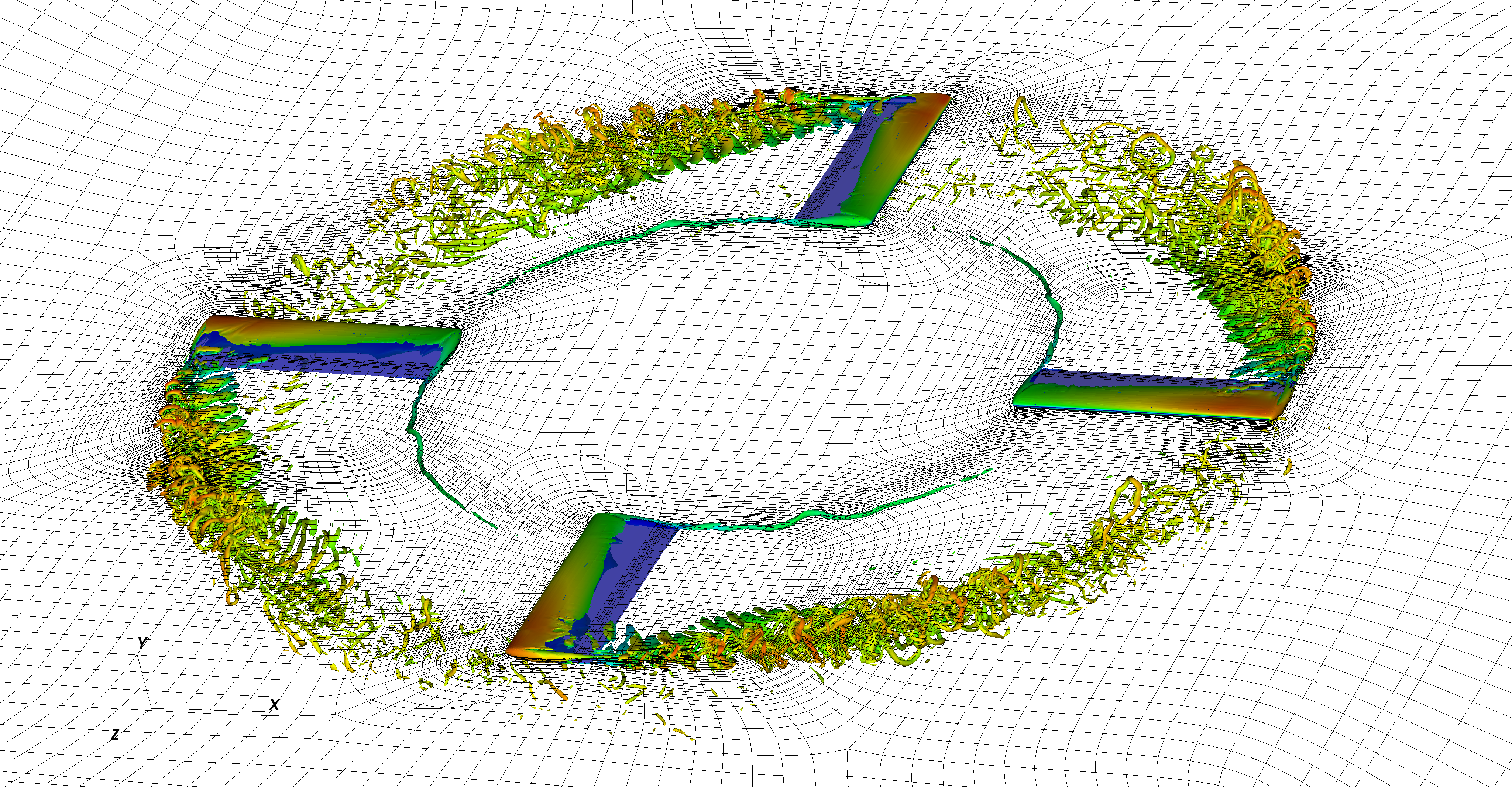} 
    \end{tabular}
    
    \caption{Flow around a toy rotor at $\rey=10000$ in a rotating reference frame computed with the AMR version of Nek5000~\cite{Offermans2020,Offermans2023}. The plot presents the iso-surfaces of the  $\lambda_{2}$-criterion of the instantaneous flow field and the cut through the domain mid-plane showing the element boundaries. Variable resolution and nonconforming interfaces are clearly visible in the wake region behind the blades.}
    \label{fig:rotor_flow}
\end{figure}

The most complex flow case presented in this work is a ``toy" rotor studied within the EU project EXCELLERAT\footnote{https://www.excellerat.eu/} in cooperation with CINECA\footnote{https://www.cineca.it/en}.
The rotor was built out of four blades with NACA0012 airfoil of length~$3$ (using an airfoil chord as a unit length), rounded wing tips and an angle of attack equal to~$5^{\circ}$. 
The simulation was performed in a rotating reference frame using an AMR framework~\cite{Offermans2020,Offermans2023}. 
The simulation time unit and an angular rotation speed of the reference frame, $\Omega$, were adjusted to set a linear velocity of an external blade tip (located at radius $r_{0}=6.5$) to~$1$. This gives the rotation period $T = 2\pi / \Omega = 2\pi r_0$ equal~$40.84$.
The Reynolds number based on the chord length and the rotation speed at the position of the external blade tip was equal to $\rey=10000$. The spectral error indicator was based on the Cartesian components of the velocity field averaged over~$0.2$ simulation time units. 
At each refinement stage $10\%$ of the elements with the highest estimated computational error were refined, and the elements with an indicated error below $5.0\times10^{-7}$ were marked for coarsening. 
This process was repeated multiple times until the required resolution was reached by increasing the number of elements from~$346336$ (initial conforming mesh) up to~$1088595$. A maximum allowed mesh resolution was defined by setting the maximum refinement level to~$3$. 
The rotor was embedded in a cylindrical domain of radius~$33.2$ and extended in the vertical direction ($y$ axis in the simulation coordinate system) from~$-18.2$ to~$13.2$. 
The simulation was run for~$1.9$ full rotations before the UQ framework was executed. 
The visualization of the instantaneous flow field at a simulation time~$t=77.6$ together with the cut through the domain mid-plane showing the element boundaries is presented in \fig~\ref{fig:rotor_flow}.

\section{Results and Discussion}\label{sec:results}
In this section, we elaborate on the application and performance assessment of the proposed in-situ UQ framework. 
\sect~\ref{sec:accuracyMACF} is focused on the assessment of the accuracy and robustness of the proposed ACF model~(\ref{eq:acfModel}).
The use of this model to estimate uncertainty in the sample mean estimator~(\ref{eq:sme}) is discussed in \sect~\ref{sec:validMACF}. 
The assessment of the in-situ estimation of such uncertainties for turbulent flow simulations introduced in \sect~\ref{sec:useCases} is detailed in \sect~\ref{sec:resultsInSitu}. 
This is followed by \sect~\ref{sec:inSituPerf} where the computational performance of the algorithm is discussed.

\subsection{Accuracy and robustness of the proposed ACF model}\label{sec:accuracyMACF}
The core requirement for precisely estimating the uncertainty of an SME using the iMACF method of \sect~\ref{sec:iMACF} is to accurately model the ACF that is plugged into \eq~(\ref{eq:varMu}). 
For this purpose, we proposed the algebraic function~(\ref{eq:acfModel}), the optimality of which is motivated in this section.
For quantitative assessment of the accuracy and robustness of the proposed ACF model, two sets of time series data are employed: velocity samples of a turbulent channel flow introduced in \sect~\ref{sec:useConformal}, and synthetic samples generated from a first-order autoregressive model, AR(1). In particular, AR(1) is defined as 
\begin{equation}\label{eq:ARM1}
x_i=a x_{i-1}+b \varepsilon_i\,,\quad i=1,2,\ldots,n \,,
\end{equation}
where~$\varepsilon_i$ are uniformly-distributed as $\varepsilon_i\sim \cU[0,1]$ and $a,b\in \mathbb{R}$. 
To ensure the time series is statistically stationary, it is required that $|a|\leq 1$.
The particular values of $a=0.1$ and $b=0.9$ are considered for the analyses in the present section and \sect~\ref{sec:validMACF}. 
For any~$n$ samples, the analytical value of the variance of the expectation of~$x$ can be obtained from 
\begin{equation}\label{eq:arm_sigma}
 n\BV[\BE[x]] = b^2/(12(1-a)^2)    \,.
\end{equation}
This expression can be used to validate the values of $\hat{\sigma}^2(\hmu)$ estimated by employing the UQ approaches introduced in \sect~\ref{sec:upUQMethods}.
Both of the data sets used in the section (channel flow and AR(1)) have a sample size of~$n=10^5$.

\begin{figure}[t]
    \centering
    \subcaptionbox{Data: AR(1) defined in~\eq(\ref{eq:ARM1}) (synthetic data).}{\includegraphics[width=0.99\linewidth]{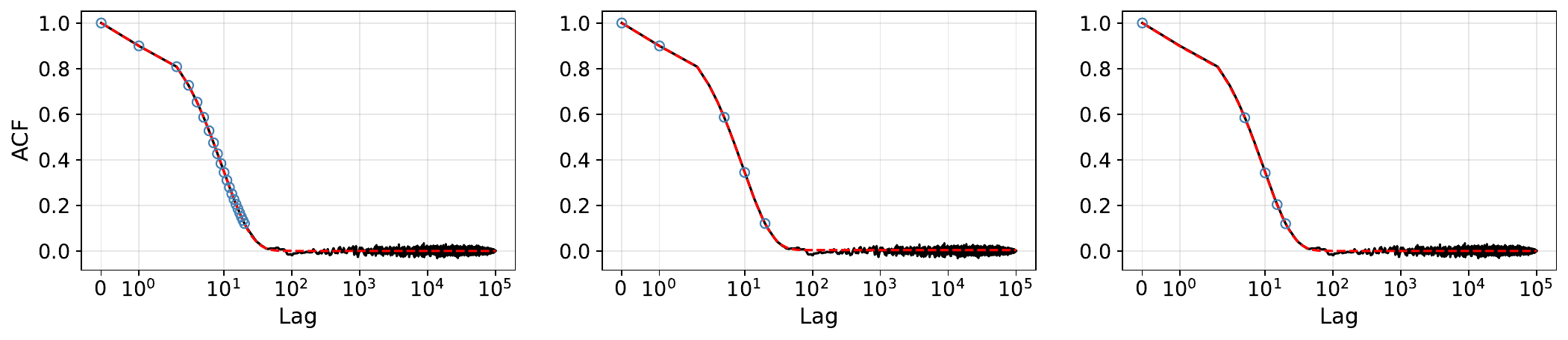}}
    \subcaptionbox{Data: $\langle u\rangle_{xz}$ at $y^+=94$ of a turbulent channel at $\reyt=300$. Note that $\langle \cdot\rangle_{xz}$ represents averaging over the wall-parallel directions~$x$ and~$z$.  
    The inner-scaled wall distance $y^+$ is defined as $y^+=y u_\tau/\nu$.}{\includegraphics[width=0.99\linewidth]{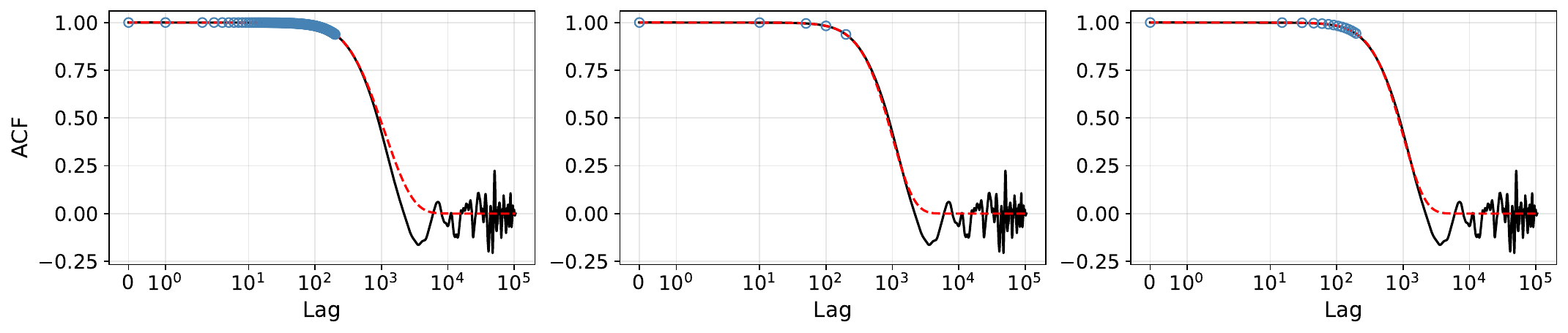}} 
    \caption{Sample estimated ACF (solid black line), sample training ACF data (markers), and modeled ACF (dashed red line) for $10^5$ samples.
    The modeled ACFs are obtained using: (left) The power-law model (see \ref{app:ARM_modeledACF}), (middle) the ACF model~(\ref{eq:acfModel}) with a sparse set of training lags $[0,1,5,10,20]$ for the synthetic data and $[0,10,50,100,200]$ for the channel flow data, and (right) the ACF model~(\ref{eq:acfModel}) with sampling intervals $5$ for the synthetic and $15$ for the channel flow data, respectively. The maximum lag at which the sample-estimated ACF data are used for training the ACF model is $20$ for the synthetic and  $200$ for the channel flow data.}
    \label{fig:fun2_toy_chan}
\end{figure}

\fig~\ref{fig:fun2_toy_chan} shows the sample-estimated (black lines) and modeled ACFs (dashed red lines) of the time series samples of the AR(1), and also the channel flow streamwise velocity samples at a distance from the wall. 
The modeled ACF is obtained by the proposed MACF model~(\ref{eq:acfModel}) using different sets of the training sample-estimated ACFs (shown by markers) which are subsets of the available samples of each of the two time series data sets.
The training sample-estimated ACFs in \fig~\ref{fig:fun2_toy_chan} are particularly chosen in two different ways over the range of lag zero and a given maximum lag: i) at a set of sparse lags with non-equal spacings (middle column), and ii) a set of lags with a fixed spacing that is a multiplication of the time lag between the original samples (right column). 
For reference, the modeled ACF from a power-law model, see \ref{app:ARM_modeledACF}, is also shown (left), which is computed using the training sample ACFs at all the original sampled times. 
For all the cases, the modeled ACF is found to be accurate for both AR(1) and channel flow data up to the point where the sample-estimated ACFs show spurious wiggles.
This particularly confirms the robustness and accuracy of the ACF model~(\ref{eq:acfModel}), since even with a small number of training samples an accurate smooth prediction of the ACF is achieved.

As explained in \sect~\ref{sec:iMACF}, for the in-situ MACF estimator (iMACF), the size of the training data must be as small as possible and at fixed sampling intervals to ensure the low-storage (memory-efficient) property.
Furthermore, the training samples should be taken at the lowest possible lags.
These requirements are met in the construction represented in the right column of \fig~\ref{fig:fun2_toy_chan}, which is also used in practice in the in-situ UQ framework as detailed in \sects~\ref{sec:resultsInSitu} and~\ref{sec:inSituPerf}.

A general rule found in the present study is to choose the sampling frequency~$T_s$ for selection of the training data such that the first non-zero time lag does not exceed~$\approx 15$ to obtain an accurate ACF model.
Less important is the maximum training lag (\ie~$M_{\max} T_s$ in \sect~\ref{sec:iMACF}), which could be chosen to minimize the number of elements in the buffer by considering the training ACF up to $\approx 0.8$. This value was found through experimentation to make a balance between the size of the buffer array and the accuracy of the modeled ACFs. 
In any case, this upper limit should not be so large to include the wiggles around zero ACF. 
These guidelines are important to be imposed in practice noting that driven by the physics, the characteristics of the ACF of turbulence time series depends on the quantities as well as the locations in the flow domain (more specifically at different wall-distances in wall-bounded flows).

The final point regarding \fig~\ref{fig:fun2_toy_chan} is that the most optimal scenario in terms of the number of training data required, is to use a sparse set of sample ACFs with non-uniform lag distances, \ie~the middle column in the figure. 
However, constructing an automatic procedure for updating the buffer in the in-situ algorithm, as it is made with a fixed sampling interval~$T_s$ in \fig~\ref{fig:iMACF}, may not be possible. 
Another interesting observation is that the ACF modeled by the power-law method (see \ref{app:ARM_modeledACF}), which is more involved than the proposed MACF, may deviate more from the sample ACF and thus result in less accurate estimates for uncertainties, see the plots in the bottom of \fig~\ref{fig:fun2_toy_chan}.

\begin{figure}[t]
    \centering
    \begin{tabular}{cc}
         \includegraphics[scale=0.43]{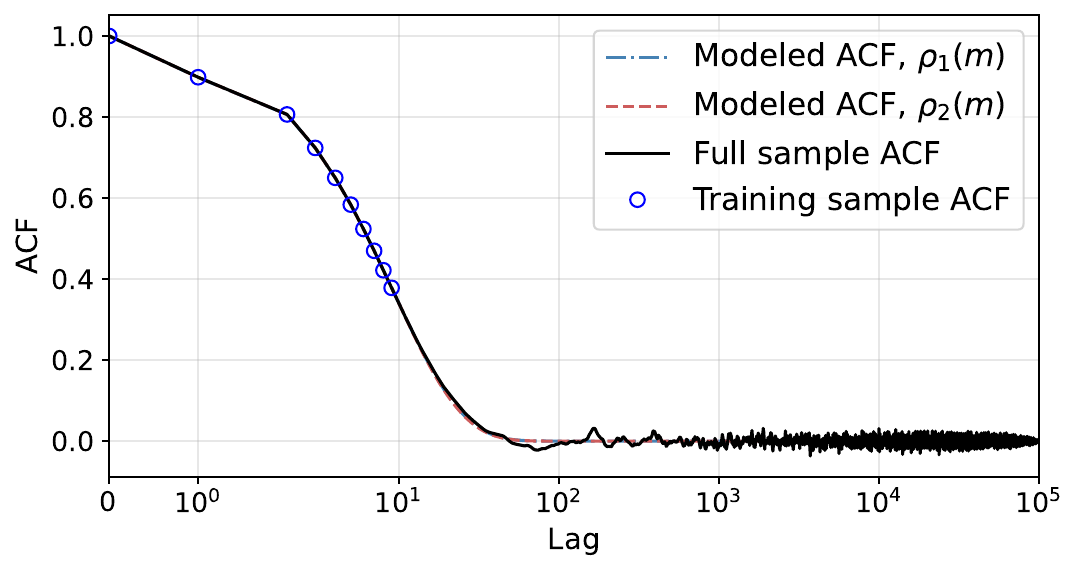} &
         \includegraphics[scale=0.43]{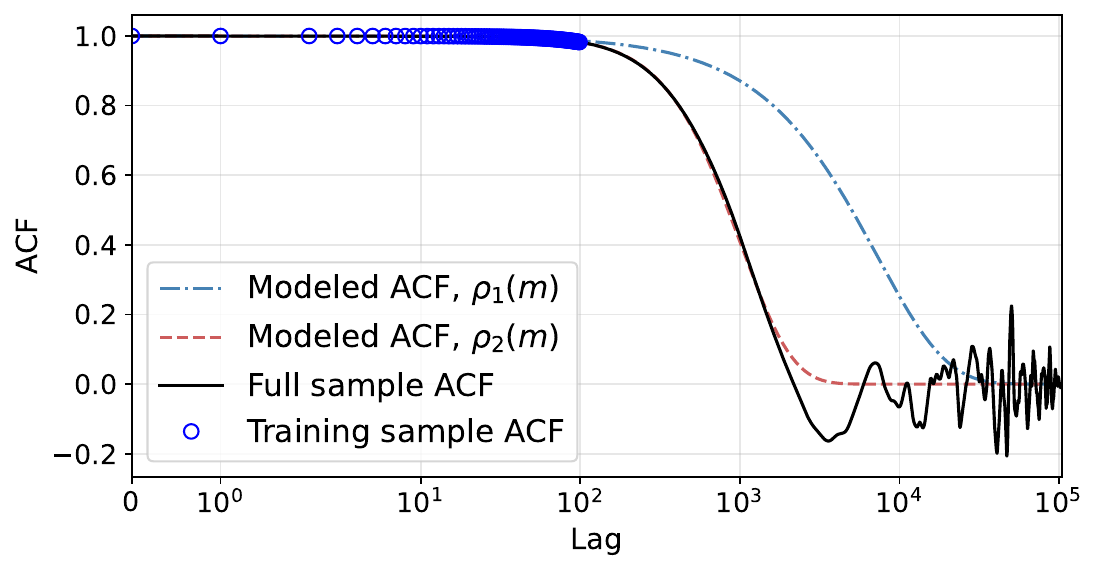} \\
    \end{tabular}
    \caption{Impact of the function used for modeling the ACF from $10^5$ time samples taken from (left) ARM(1) defined in~\eq~(\ref{eq:ARM1}), and (right) the turbulent channel flow averaged streamwise velocity~$\langle u\rangle_{xz}$ at $y^+=94$. The training sample-estimated ACF data are taken at the first~$10$ lags for the AR(1) and $100$ lags for the channel flow data, respectively. The ACF models $\rho_1(m)$ and $\rho_2(m)$ refer to \eqs~(\ref{eq:acfModel}) and (\ref{eq:MACF_fun1}), respectively.}
    \label{fig:fun1_fun2_toy_channel}
\end{figure}

As briefly mentioned in \sect~\ref{sec:uqOverview}, in some previous studies in the literature, function~(\ref{eq:MACF_fun1}) has been used to model the ACF. 
However, \fig~\ref{fig:fun1_fun2_toy_channel} shows that the use of this model is limited to processes such as AR(1), where the integral time scale, \ie~the area below the ACF-time curve, is small. 
Conversely, for the turbulence time series where the history effects last for a longer time and the corresponding autoregressive model is of high order, the ACF model~(\ref{eq:MACF_fun1}) fails to accurately obtain a smooth function for the ACF.
In particular, in \fig~\ref{fig:fun1_fun2_toy_channel}, the ACF inaccurately modeled by \eq~(\ref{eq:MACF_fun1}) is extended over longer time lags compared to the sample-estimated ACF, resulting in an over-estimation of the uncertainty in the associated SME when it is employed in \eq~(\ref{eq:varMu}).

\subsection{Validation and robustness of the uncertainties estimated by the MACF method}\label{sec:validMACF}
As motivated above, function~(\ref{eq:acfModel}) can lead to accurate smooth models for the ACF, which can be plugged into \eq~(\ref{eq:varMu}) to accurately estimate~$\hsig(\hmu)$, \ie~the standard-deviation of a sample mean estimation~$\hmu$. 
This will be shown in this section with two different studies applied to the data of the AR(1) defined in \eq~(\ref{eq:ARM1}), and the turbulent channel flow velocity data.

For the first study, several realizations of a time series are generated, and the distribution of their sample-estimated SME's uncertainty is compared to the associated empirical uncertainty obtained from \eq~(\ref{eq:empiricStd}). The latter can reflect the true or population uncertainty. 
For this purpose, an ensemble of size~$N_e$ of the computationally inexpensive AR(1), \eq~(\ref{eq:ARM1}), is considered. 
Here, each of the AR(1) simulations starts from an independent random sample taken from the uniform distribution $\mathcal{U}[0,1]$ and continues to~$1.5 n$ samples. 
The first~$0.5 n$ samples are discarded to account for the burn-in. 
Hence, for each of the~$N_e$ realizations there are~$n$ samples of the AR(1) to which the MACF estimator is applied for estimating the corresponding~$\hsig(\hmu)$. 
For the MACF estimator to be consistent, the probability density function (PDF) of the resulting~$\hsig(\hmu)$ should contain the empirically-estimated uncertainty of the ensemble expectation~$\mu_e$,
 \begin{equation}\label{eq:empiricMean}
     \mu_e=\mathbb{E}[\hat{\mu}] \approx \frac{1}{N_e}\sum_{i=1}^{N_e} \hat{\mu}_i  \,, 
 \end{equation}
measured by~$\sigma_e$ that is given by, 
\begin{equation}\label{eq:empiricStd} 
\sigma^2_e=\mathbb{V}[\hat{\mu}] \approx \frac{1}{N_e}\sum_{i=1}^{N_e} \left(\hat{\mu}_i - \mu_e \right)^2 \,.
 \end{equation}
 
\begin{figure}[t]
    \centering
    \begin{tabular}{ccc}
         \includegraphics[scale=0.52]{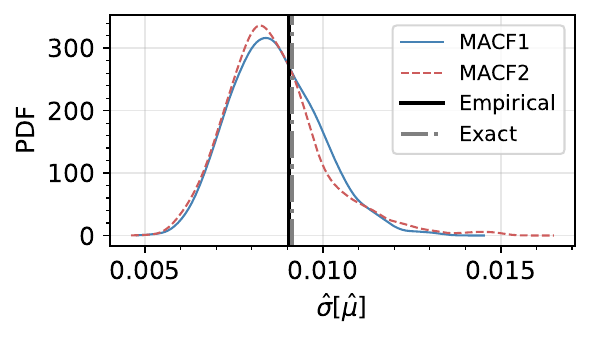} \hspace{-0.5cm} &
         \includegraphics[scale=0.52]{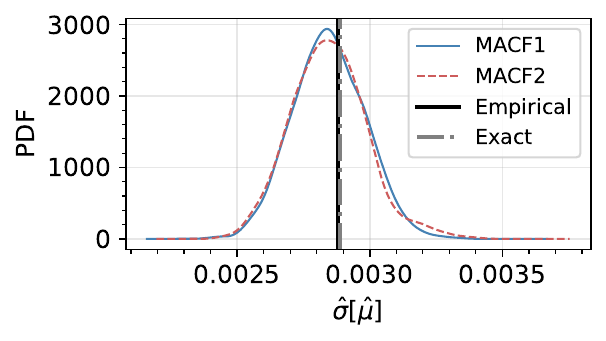} \hspace{-0.6cm} &
         \includegraphics[scale=0.52]{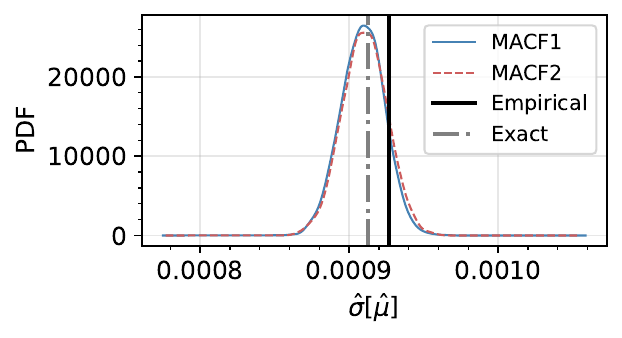}
    \end{tabular}
    \caption{Validation of the $\hat{\sigma}(\hat{\mu})$ estimated by the MACF method using the samples of the ARM(1) defined in \eq~(\ref{eq:ARM1}). The plots show the PDF of the~$\hat{\sigma}(\hat{\mu})$ using the sample-estimated ACF at the first time lags (MACF1) and lags $[0,2,5,8,15]$ (MACF2) for training the ACF model~(\ref{eq:acfModel}). The PDFs are obtained by $2000$ repetitions of the independent realizations of the AR(1) with the sample size~$n$ per realization being equal to (left) $10^3$, (middle) $10^4$, and (right) $10^5$. The solid and dash-dotted vertical lines respectively show the empirical estimate~$\sigma_e$ given by \eq~(\ref{eq:empiricStd}), and the exact value of~$\sigma(\hmu)$ obtained from~\eq~(\ref{eq:arm_sigma}).}
    \label{fig:pdfMACF_toy}
\end{figure}

\fig~\ref{fig:pdfMACF_toy} illustrates the outcome of the described procedure for different values of the sample size~$n$ where the ACF modeled by \eq~(\ref{eq:acfModel}) is constructed using two different sets of training sample-estimated ACFs.
In all cases, the empirical~$\sigma_e$ is very close to the exact~$\sigma(\hmu)$ (given by \eq~(\ref{eq:arm_sigma})), and falls within the PDF of the~$\hsig(\hmu)$ estimated by the MACF method. 
Note that as expected, by increasing the sample size~$n$ the PDF of~$\hsig(\hmu)$ becomes narrower, but more importantly the mode of the PDF (most probable value of~$\hsig(\hmu)$) becomes almost the same as the exact~$\sigma(\hmu)$.

\begin{figure}[t]
    \centering
    \includegraphics[scale=0.5]{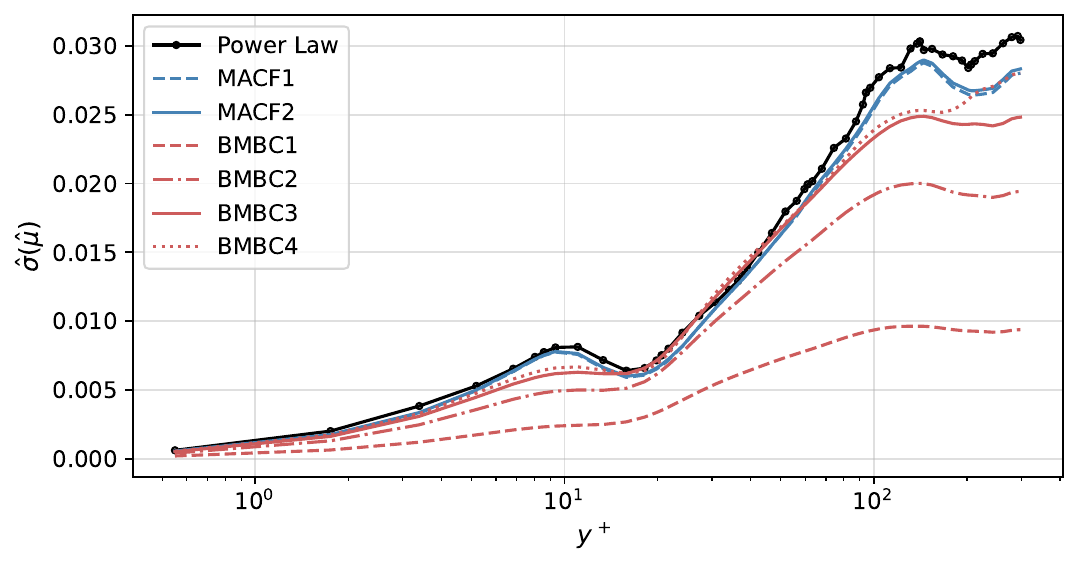}
    \caption{Estimated standard deviation of the SME of $\langle u \rangle_{xz}$, the averaged streamwise velocity of the turbulent channel flow, versus the inner-scaled wall distance~$y^+$ ($10^5$ time samples are used). The following hyperparameters are used for different uncertainty estimators; power-law method (see \ref{app:ARM_modeledACF}): the order of the ARM is $p=100$ and the sample-estimated ACFs at first 300 time lags are used for training the ACF model; MACF1 and MACF2: the MACF method with the ACF model~(\ref{eq:acfModel}) trained by the sample-estimated ACF at first 300 lags (MACF1) and lags corresponding to [0,10,50,100,200,250,300] (MACF2); BMBC1 to BMBC4: the BMBC method with the batch size equals to 100, 500, 1000, 2000, respectively.
    }\label{fig:chan_uncertCompareEstims}
\end{figure}

For a single realization of the turbulent channel flow, a study to validate the~$\hsig(\hmu)$ estimated by the MACF method is to make a comparison with the power-law method which has been used in the literature as in \rfs~\cite{oliver:14,xavier:21}, see \ref{app:ARM_modeledACF}.
According to \fig~\ref{fig:chan_uncertCompareEstims}, at all wall-normal locations, the agreement between the MACF and power-law methods is good, in general. 
There is a small discrepancy in the outer layer of the mean velocity profile, \ie~$y^+\gtrsim 150$, which originates from the slight imperfection of the modeled ACF potentially by both methods (although in \fig~\ref{fig:fun2_toy_chan}, the power-law method is found to be slightly less accurate than MACF for modeling the ACF). 
To ensure the robustness of the MACF method, the estimated uncertainties associated to two different sets of ACF training samples are found to be almost the same.
This is in fact a main advantage of the proposed ACF model~(\ref{eq:acfModel}) for estimating~$\hsig(\hmu)$, \ie~the resulting estimates are not biased with respect to the choice of the hyperparameters. 
To make this point clearer, plots from the BMBC approach~\cite{russo:17} are also provided in \fig~\ref{fig:chan_uncertCompareEstims}. 
Clearly, the estimated~$\hsig(\hmu)$ are sensitive to the batch-size and for none of the considered batch sizes the estimates would be close enough to those by the MACF and power-law methods. 
As formerly stated, this potential bias in the~$\hsig(\hmu)$ values is the main barrier for using the batch-based methods for in-situ applications. 
But if we need to choose one of such approaches, the BMBC approach of~\cite{russo:17} is preferred to the NOBM, see the discussion in~\ref{app:batchEffect}.

\subsection{Robustness of the in-situ UQ algorithm}\label{sec:resultsInSitu}
In this section, we consider the iMACF algorithm~\ref{alg:iMACF}, the in-situ version of the MACF method described in \sect~\ref{sec:iMACF}. 
The aim is to assess the impact of the relevant hyperparameters on the accuracy of the modeled ACF and the resulting~$\hsig({\hmu})$ when the sample ACFs are computed in an in-situ way following \eq~(\ref{eq:upACF}).
To this end, the time series of the streamwise velocity of the turbulent channel flow at any distance from the wall can be considered. 
The three hyperparameters of the iMACF method investigated here are~$m_{\max}$, the maximum lag up to which the sample-estimated ACF from \eqs~(\ref{eq:acov_op}) and~(\ref{eq:upACF}) are used to train the ACF model~(\ref{eq:acfModel}), $T_s$, the sampling interval, and, $n$ the sample size. 

\begin{figure}
    \centering
    \begin{tabular}{cc}
         \includegraphics[scale=0.4]{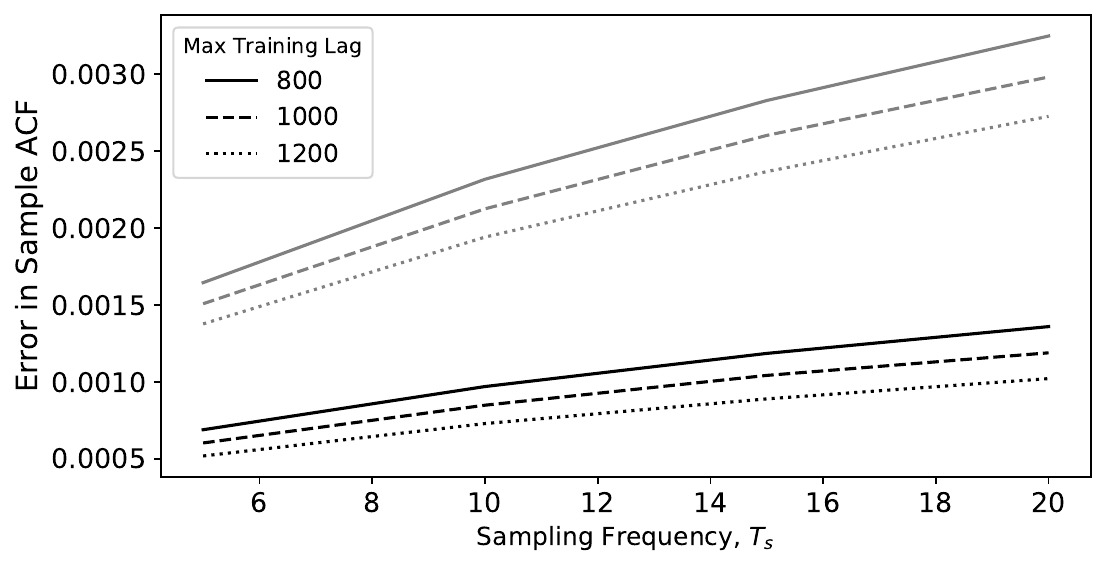} & 
         \includegraphics[scale=0.4]{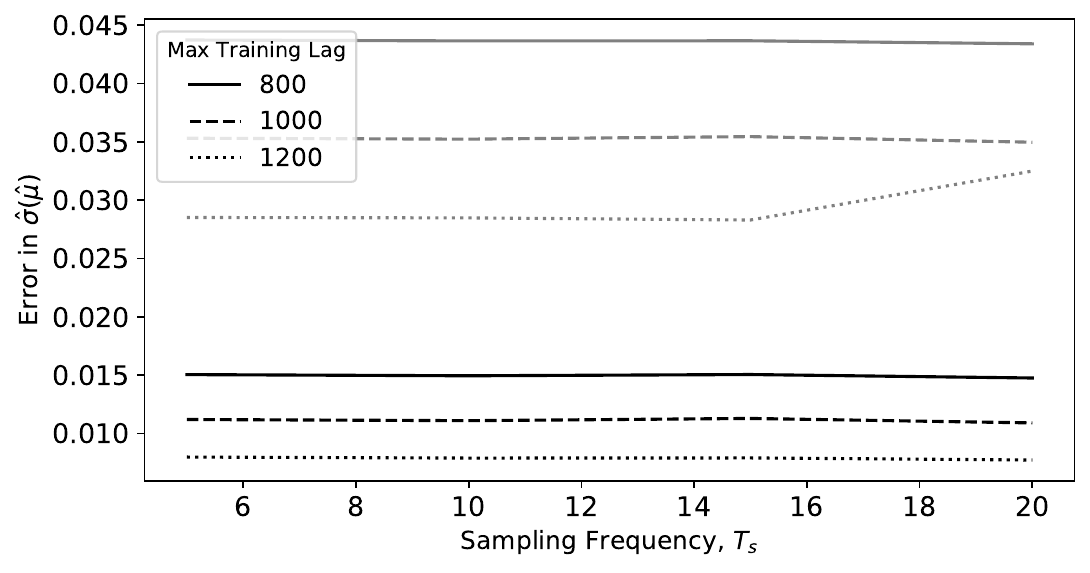} \\      
         {\small (a)} & {\small (b)} \\
    \end{tabular}
    \caption{Impact of the maximum training lag~$m_{\max}$, sampling frequency~$T_s$ and sample size~$n$ on the error in (a) the sample-estimated ACF, $e_\rho$, and (b) the standard deviation of the SME of $\langle u \rangle$, $e_{\hsig}$, obtained from the iMACF method. The data belong to the channel flow streamwise velocity~$\langle u\rangle_{xz}$ at the wall distance $y^+=262$ with size $n$ equal (black lines) $100000$ and (gray lines) $50000$. The error in the plots (a) and (b) is defined by \eqs~(\ref{eq:e_rho}) and (\ref{eq:e_sig}), respectively.}
    \label{fig:iMACF_hyperParam_chan}
\end{figure}

The two metrics evaluated and plotted in \fig~\ref{fig:iMACF_hyperParam_chan} are the error in the sample-estimated ACF,~$e_\rho$, (left plot) 
and the error in estimated~$\hsig(\hmu)$, $e_{\hsig}$, (right plot) that are respectively defined as,
\begin{equation}\label{eq:e_rho}
e_\rho=\|\rho_{st} (m) - \rho_{up} (m)\|_2  / M_{\max} \,,
\end{equation}
\begin{equation}\label{eq:e_sig}
e_{\hsig}=|\hsig_{st}(\hmu) - \hsig_{up} (\hmu)|/\hsig_{st}(\hmu) \,.
\end{equation}
The subscripts~\emph{st} and~\emph{up} denote the standard (offline using all data) and updating (in-situ) values, respectively. 
When evaluating~$e_\rho$, the~$M_{\max}$ is set as $M_{\max}=\text{int}(m_{\max}/T_s)$.
According to \fig~\ref{fig:iMACF_hyperParam_chan}, for fixed values of~$n$ and~$m_{\max}$, less frequent sampling (\ie~higher~$T_s$) leads to higher values of~$e_\rho$, however, the corresponding impact on the estimated~$e_{\hsig}$ is negligible. 
What, instead, has the highest influence on~$e_{\hsig}$ is the variation of~$n$ and~$M_{\max}$.
As the number of samples decreases, meaning that the averaging is run for a shorter time interval, both~$e_\rho$ and~$e_{\hsig}$ increase. 
Considering more lags for modeling an ACF, \ie~increasing~$M_{\max}$, can lead to more accurate ACF computed in an updating fashion and consequently more accurate~$\hsig$. 
In any case, as shown for the particular data set here, the largest error in~$\hsig$ estimated by the updating algorithm for the considered range of hyperparameters is less than~$5\%$ compared to what could be estimated by the standard method. 
This confirms the robustness and accuracy of the iMACF algorithm~\ref{alg:iMACF}.

\subsection{Application to complex flows}\label{sec:complFlows}
In this section, we study the application of the in-situ UQ framework to the scale-resolving simulations of the turbulent flow around a NACA4412 wing section and a moving rotor with an adaptively refined mesh as introduced in \sects~\ref{sec:useConformal} and~\ref{sec:AMRrotor}, respectively.
In the former case, we compare the iMACF approach with an incremental implementation of the BMBC method for three different batch sizes, while the latter demonstrates the integration of the in-situ UQ framework with a transient AMR simulation. In both cases, the iMACF approach is applied to all points of the three-dimensional computational grid. 
As explained in \sect~\ref{sec:iMACF}, in the iMACF algorithm~\ref{alg:iMACF} we only run lines 7 to 13 during the simulation while line 14 is executed offline. 
There are multiple reasons for this choice. First, the fitting of the ACF model to the training data in line 14 usually contributes most to the overall computing time of the UQ workflow and can thus extend the total simulation time. However, there is no need to do this step online unless we want to monitor the uncertainty of the turbulence statistics during the runtime. Second, the user might only be interested in the UQ results for a specific region of the simulation domain, such as a~2D slice or a~3D subdomain. 
Selection of these can be done after the termination of the simulation to cut down the computing time significantly. Third, there are a few hyperparameters that can influence the convergence of fitting the MACF model, including the selection of training lags. 
Thus, by running Line 14 in \alg~\ref{alg:iMACF} offline we can avoid faulty results that would require a rerun of the entire CFD simulation. 
Therefore, at the end of the simulation, the ACF training values are saved on disk, where the values have been updated during the simulation using the streaming algorithm. 
Depending on the maximum number of the training lags,~$M_{\max}$, the size of data to be stored is equivalent to only a couple of flow snapshots, and hence a small fraction of the data processed during the in-situ iMACF algorithm. 

The NACA4412 use case is run for overall~$5.37$ time units equivalent to~$200000$ time steps. 
For the UQ estimate of the streamwise component of the velocity, $u$, we apply both the iMACF and iBMBC methods considering every~$20$-th time sample of the simulation (\ie~$T_s=20$). 
For the iMACF algorithm~\ref{alg:iMACF}, we choose~$M_{\max}=50$ to ensure sample ACF values are computed over a long enough range of lags at all spatial locations within the region of interest. 
However, to improve the quality of the modeled ACF especially at larger lags~$m$, only a selection of the training ACFs was used by i) neglecting lags with the ACF values below~$\approx 0.4$ and ii) using ACFs at every $5$-th lag. Both decisions are drawn from what was observed from the a-priori tests in \fig~\ref{fig:fun2_toy_chan} (bottom row, left and middle). Altogether, the UQ results only show weak dependence on these parameters.
It is noteworthy that we did not use a larger~$T_s$ during the in-situ process to avoid increasing the uncertainties in the updating SMEs and also due to the fact that the first non-zero lag at which the sample ACF is computed should not be too large to avoid errors in the modeled ACF, see the discussion in \sect~{\ref{sec:accuracyMACF}}.

\begin{figure}[t]
    \centering
    \begin{tabular}{cc}
        \includegraphics[scale=0.103]{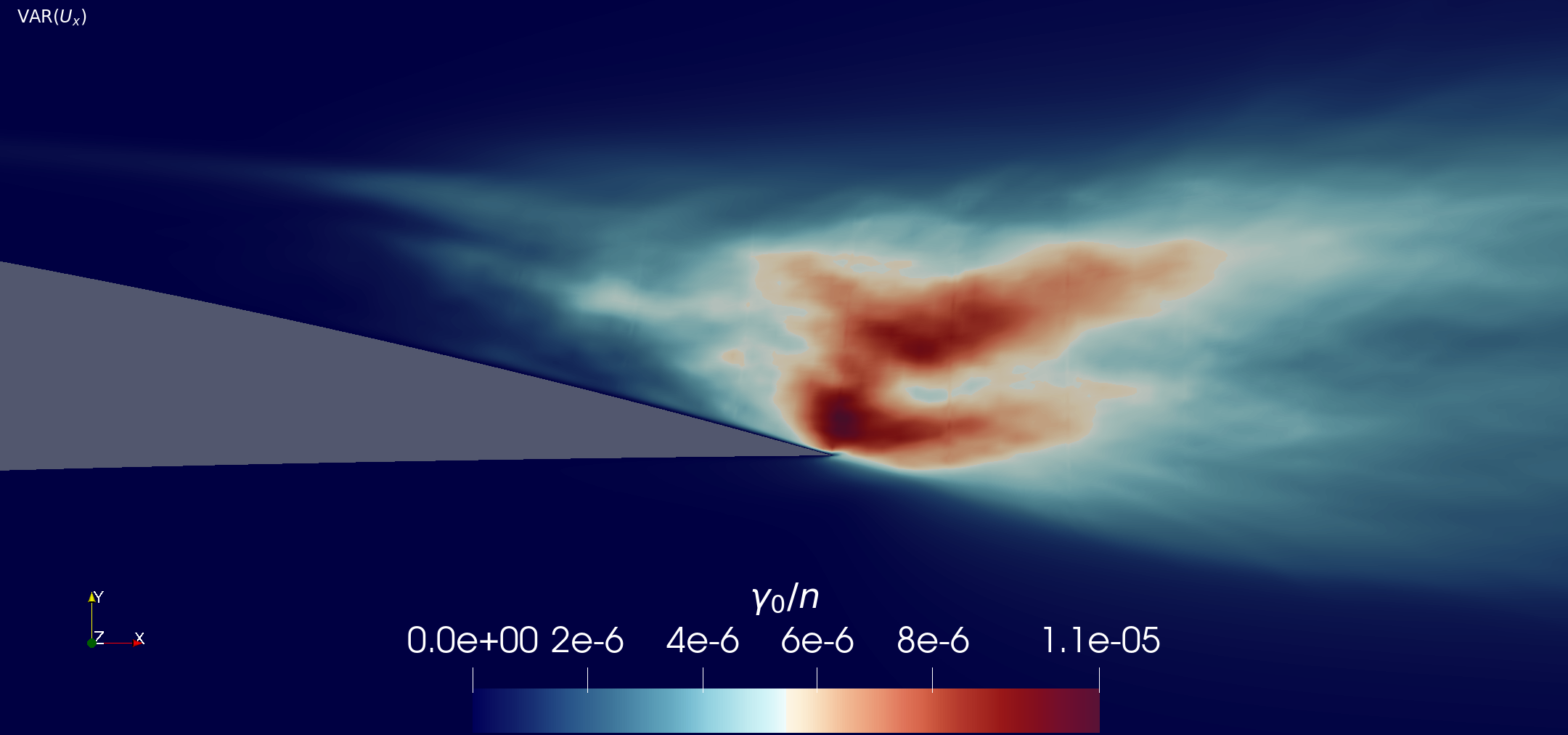} &
        \includegraphics[scale=0.103]{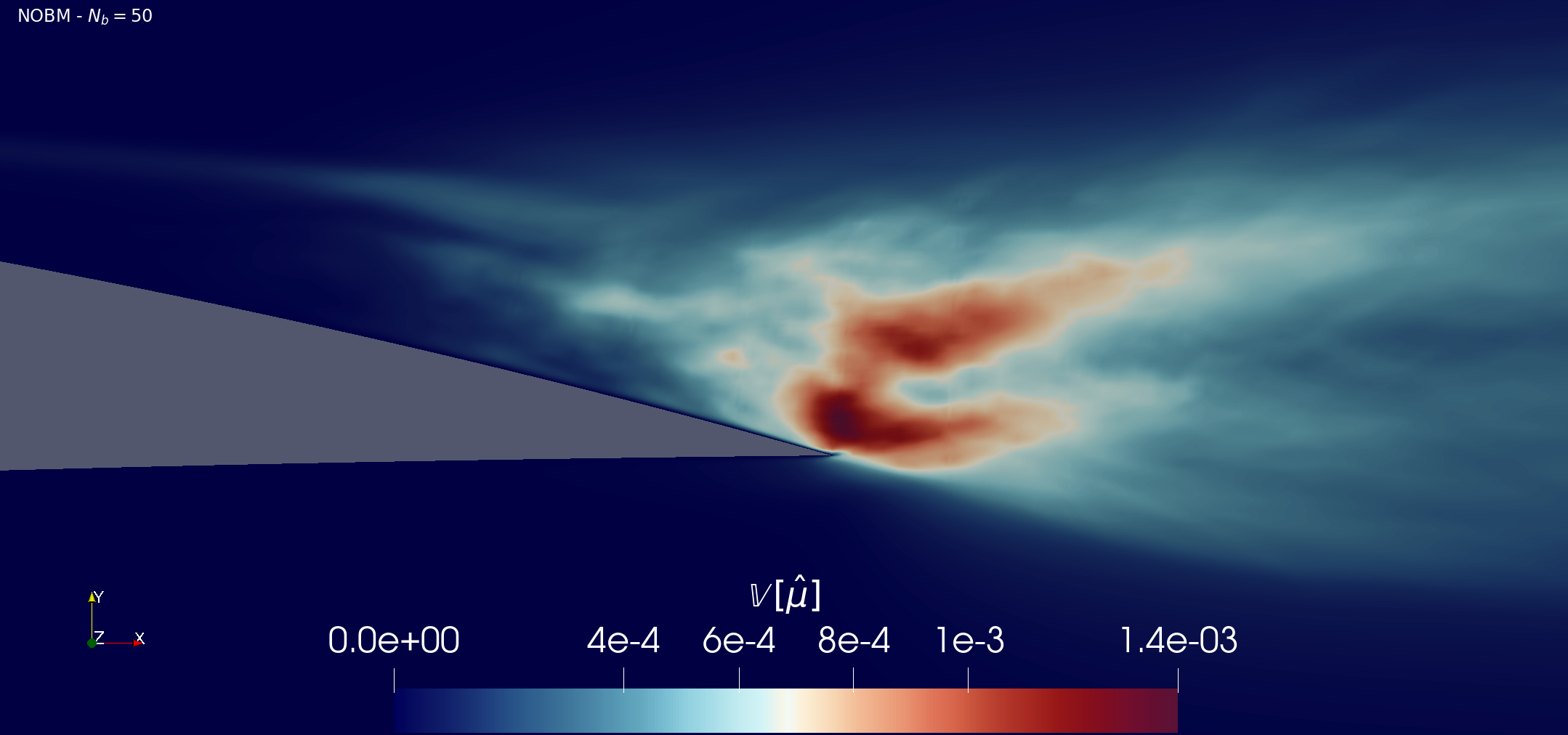} \\
        {\small (a)} & {\small (b)}\\
        \includegraphics[scale=0.103]{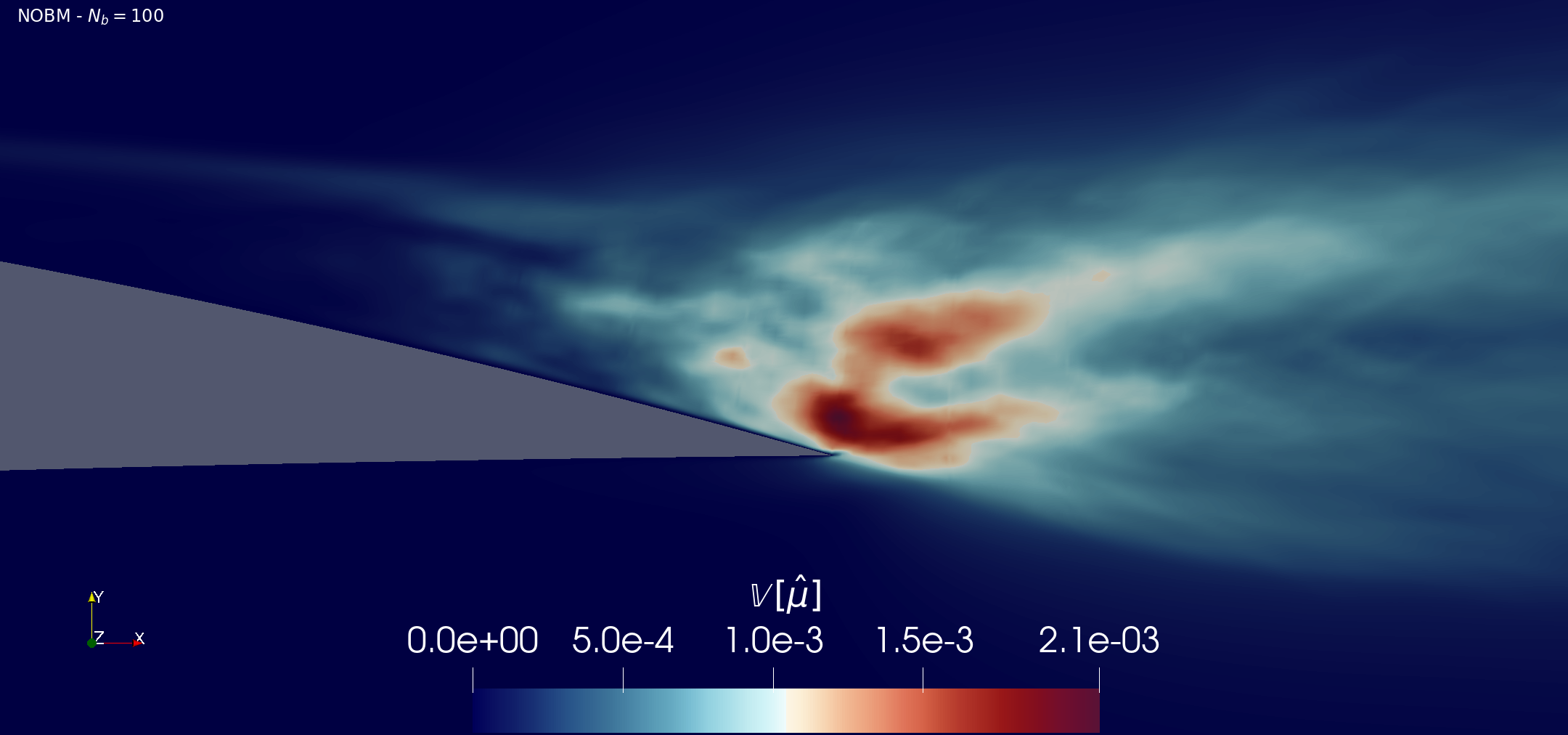} &
        \includegraphics[scale=0.103]{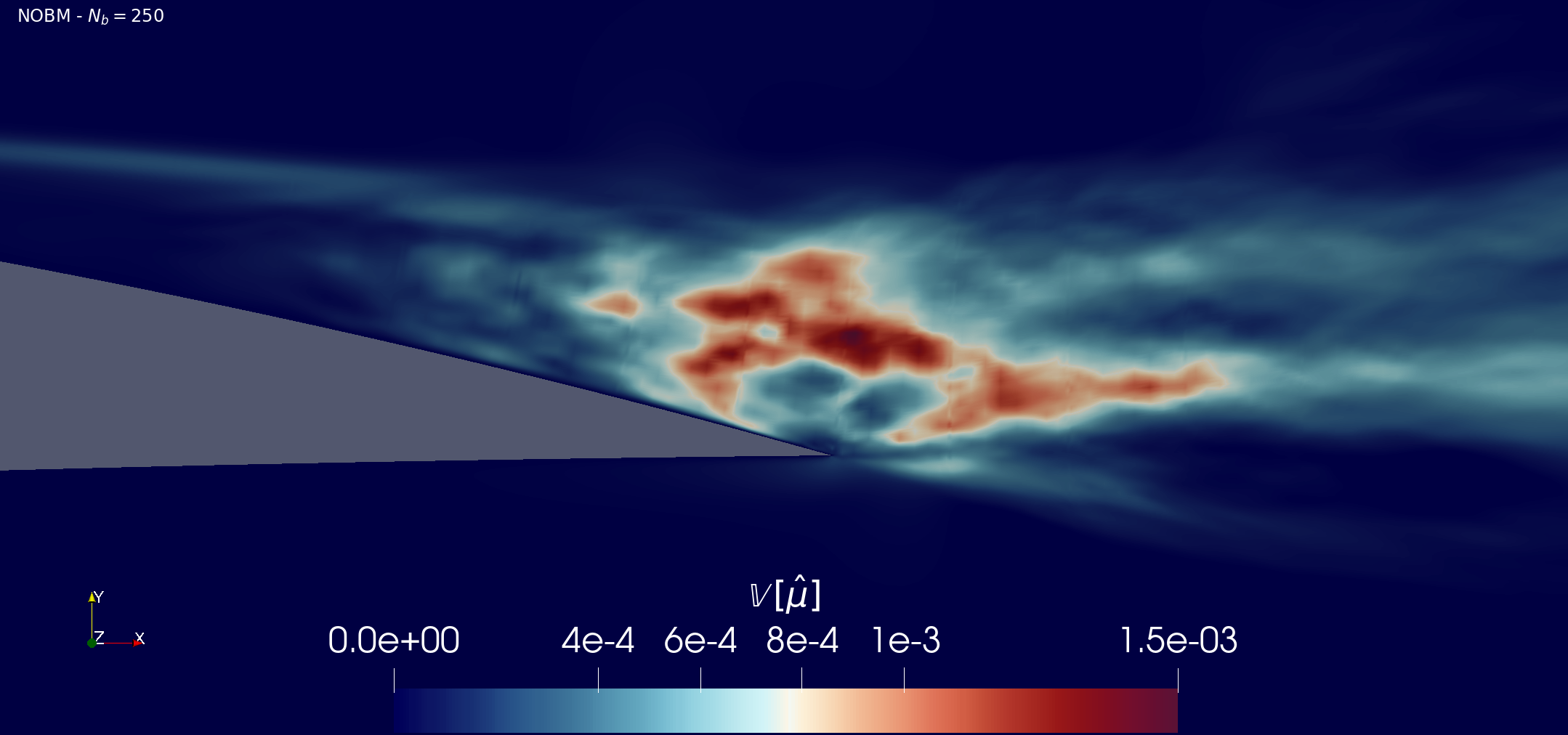}\\
        {\small (c)} & {\small (d)}\\
    \end{tabular}
    \caption{Contours of the (a) sample variance of the streamwise velocity component $u$ of the NACA4412 wing simulation at $z=0.025$ plane, and (b)-(d) estimated variance of associated SME using the iBMBC method with the batch size equal to (b) 50, (c) 100, and (d) 250.}
    \label{fig:naca4412_bmbc}
\end{figure}

\begin{figure}[t]
    \centering
    \begin{tabular}{cc}
        \includegraphics[scale=0.103]{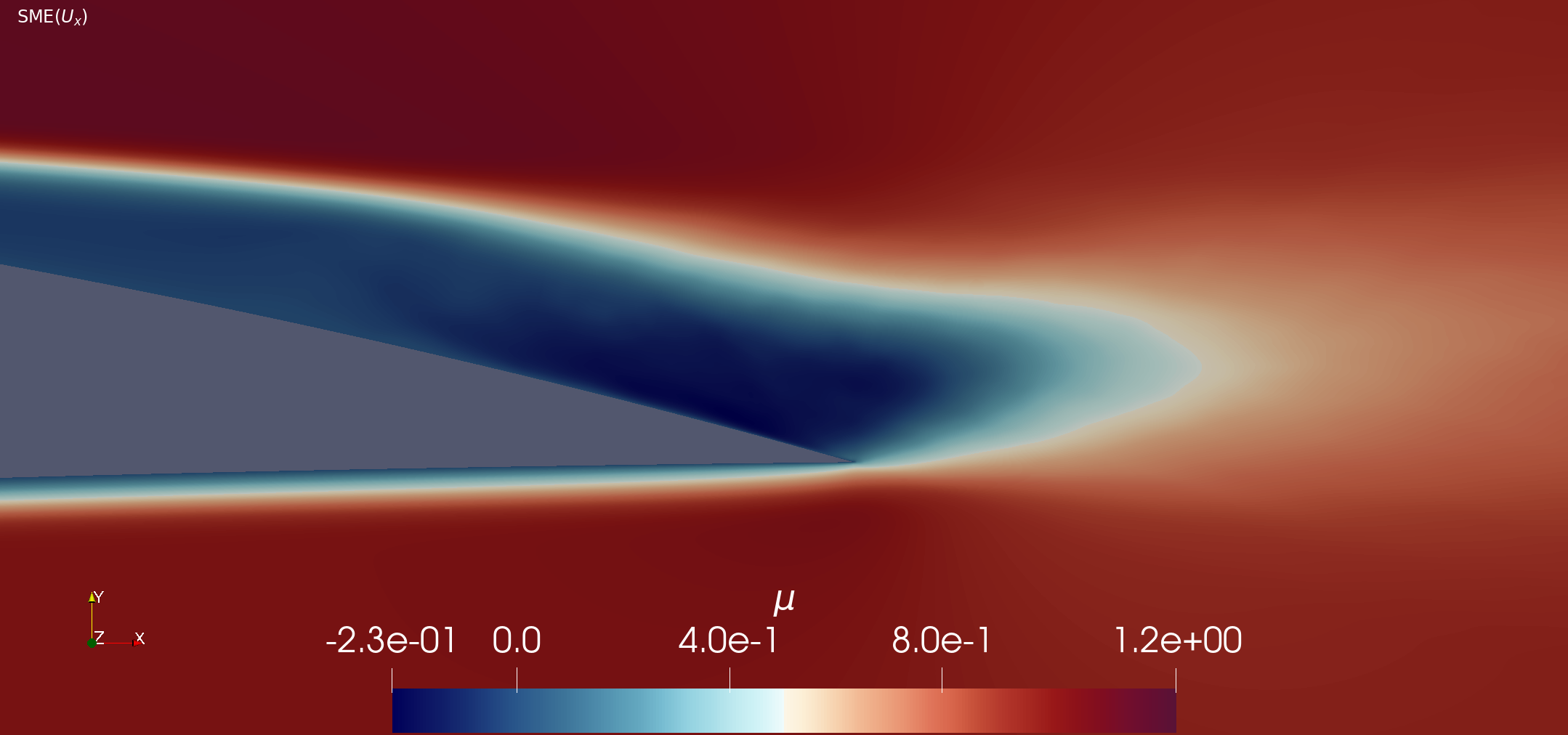} &
        \includegraphics[scale=0.103]{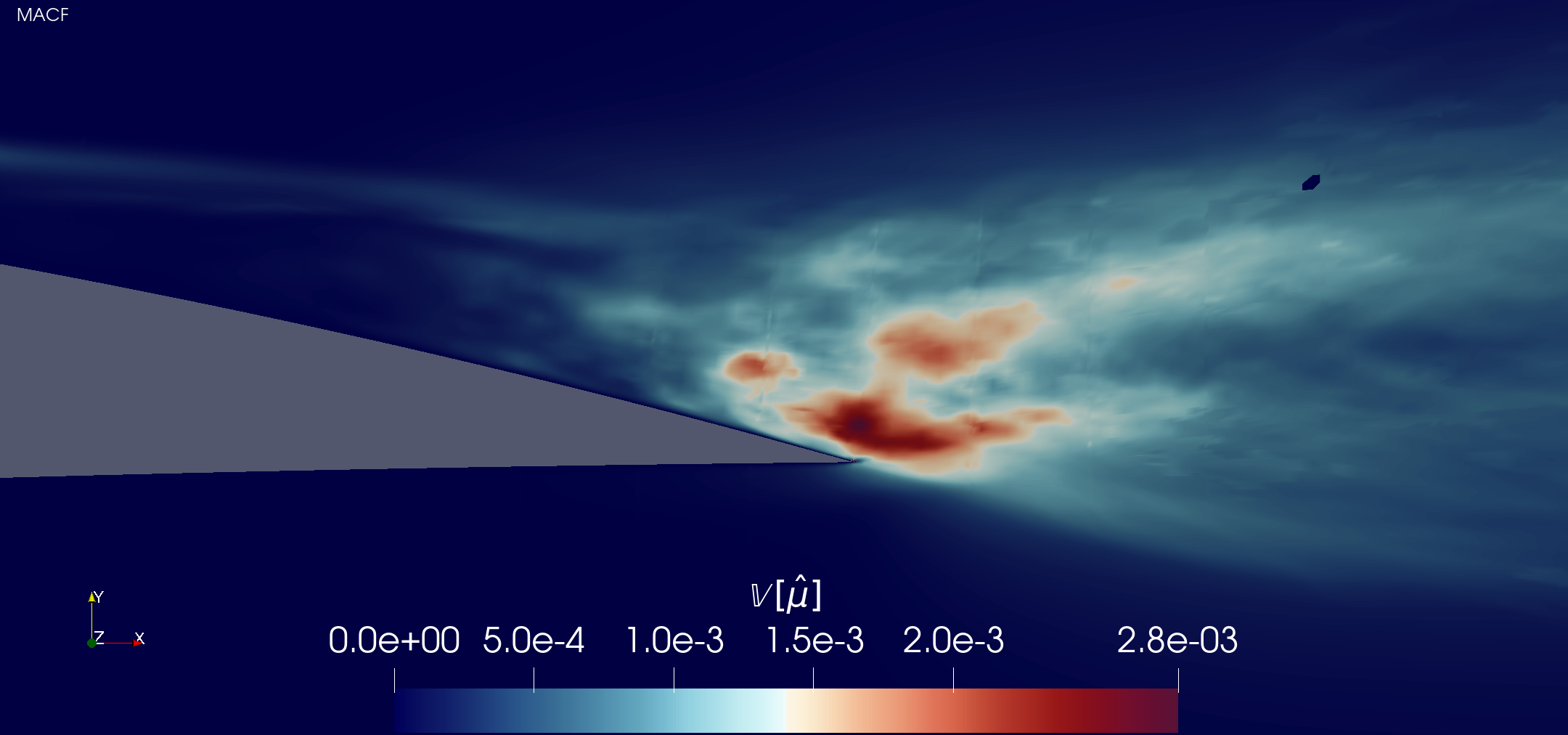}\\
        {\small (a)} & {\small (b)}\\
    \end{tabular}
    \caption{Contours of (a) the SME of streamwise velocity $u$ and (b) associated variance estimated by the iMACF method for the NACA4412 wing use case.}
    \label{fig:naca4412_macf}
\end{figure}

The UQ results of applying the iBMBC and iMACF to the NACA4412 wing are shown in \figs~\ref{fig:naca4412_bmbc} and~\ref{fig:naca4412_macf}, respectively. 
The plots are made at a 2D slice through the midplane in the spanwise direction of the wing at~$z=0.025$. 
As expected, the estimated uncertainties by the BMBC method strongly depend on the batch size.
For small batch sizes, the BMBC method underestimates the uncertainty and convergences towards the sample variance where the correct impact of the autocorrelations is ignored. 
For large batches, the number of batch means falls below the limit necessary for unbiased estimation of the uncertainties. 
These findings are in accordance with \eq~(\ref{eq:varMu}): The uncertainty in an SME depends on  both the variance of the time series and the variation of the ACF with time lag. 
Thus, an underestimated uncertainty means failing in accurate estimation of these two contributors, where the role of the modeled ACF is more probable. 
We should bear in mind that the batch size imposes a cut-off to a modeled ACF, see~\cite{xavier:21}.
Therefore, there is only a small range of batch sizes that could lead to reliable uncertainty estimates, where the range depends on the spatial location and QoI. 
In fact, for the NACA4412 simulation, we observed a difference in the inflection point of the ACF of up to two orders of magnitude depending on the specific location in the flow domain. 
In conclusion, applying the BMBC method in an in-situ way cannot be automated since the proper batch sizes cannot be chosen adaptively, instead they have to be chosen a priori. 
For the batch size of~$100$ samples, the BMBC results are found to be most similar to those of the MACF approach. 
However, we still see slightly lower values of the BMBC uncertainties and small differences in their spatial distribution that can be explained by the observations described above.

\begin{figure}[t]
    \centering
    \begin{tabular}{cc}
        \includegraphics[scale=0.103]{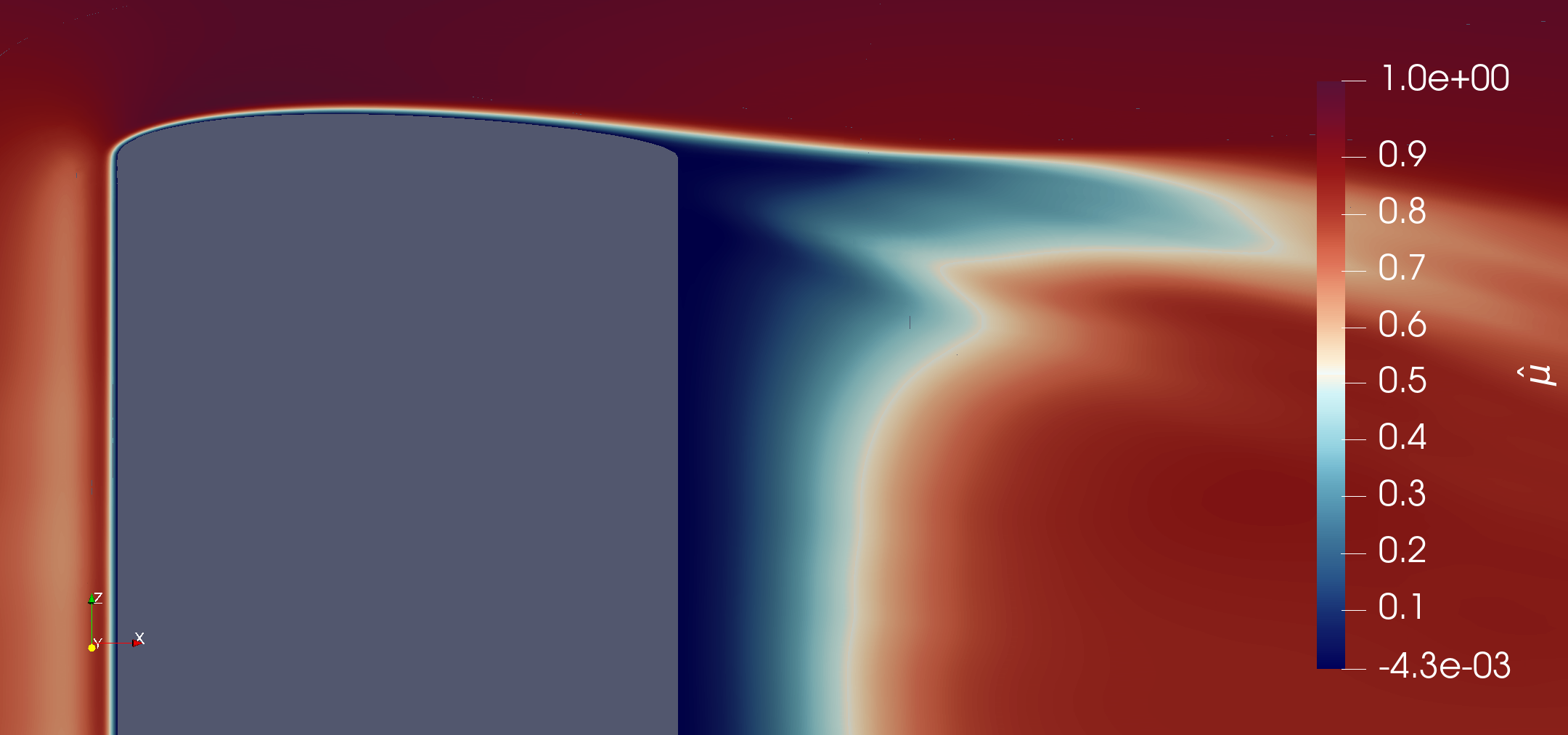} &
        \includegraphics[scale=0.103]{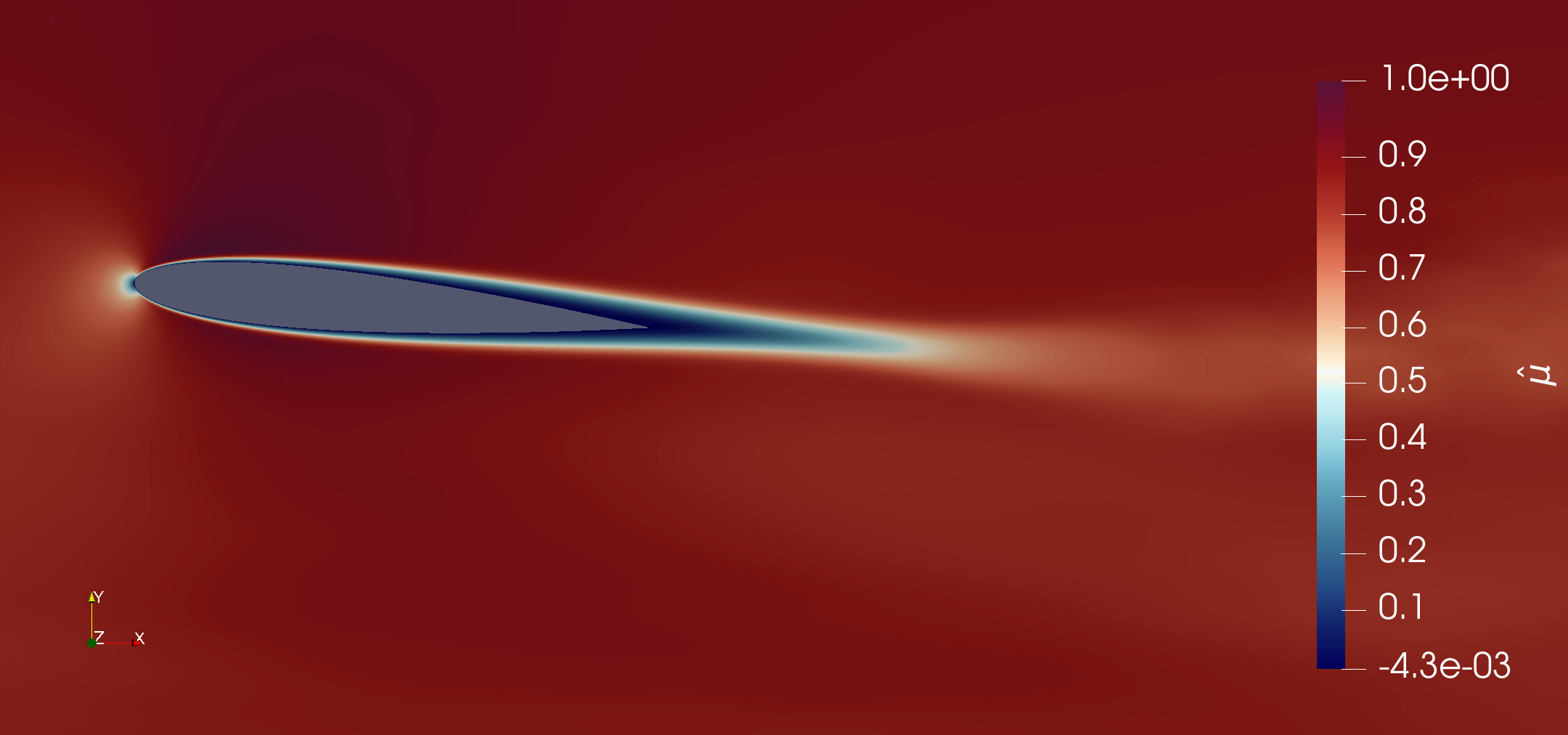} \\
        {\small (a)} & {\small (b)}\\
        \includegraphics[scale=0.103]{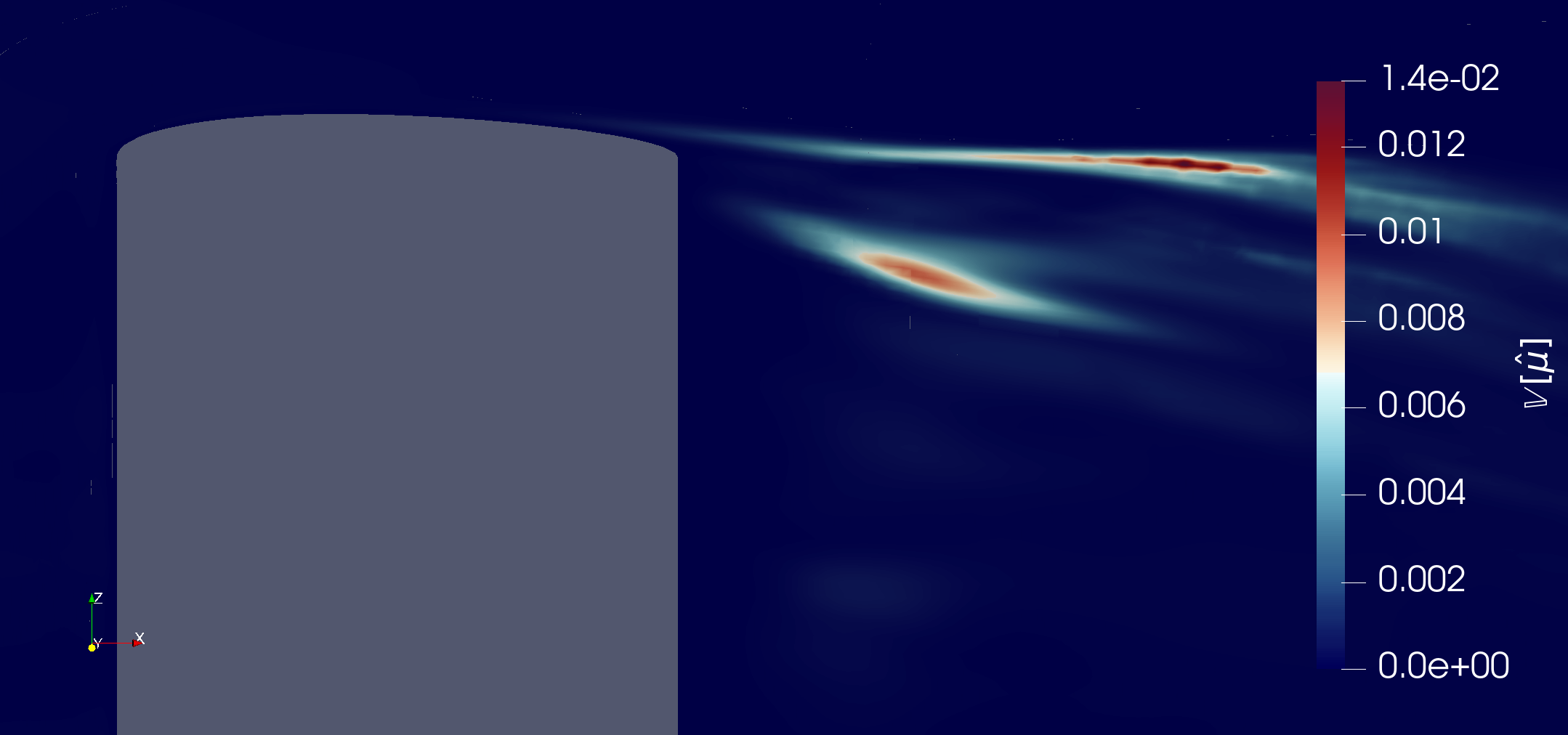} &
        \includegraphics[scale=0.103]{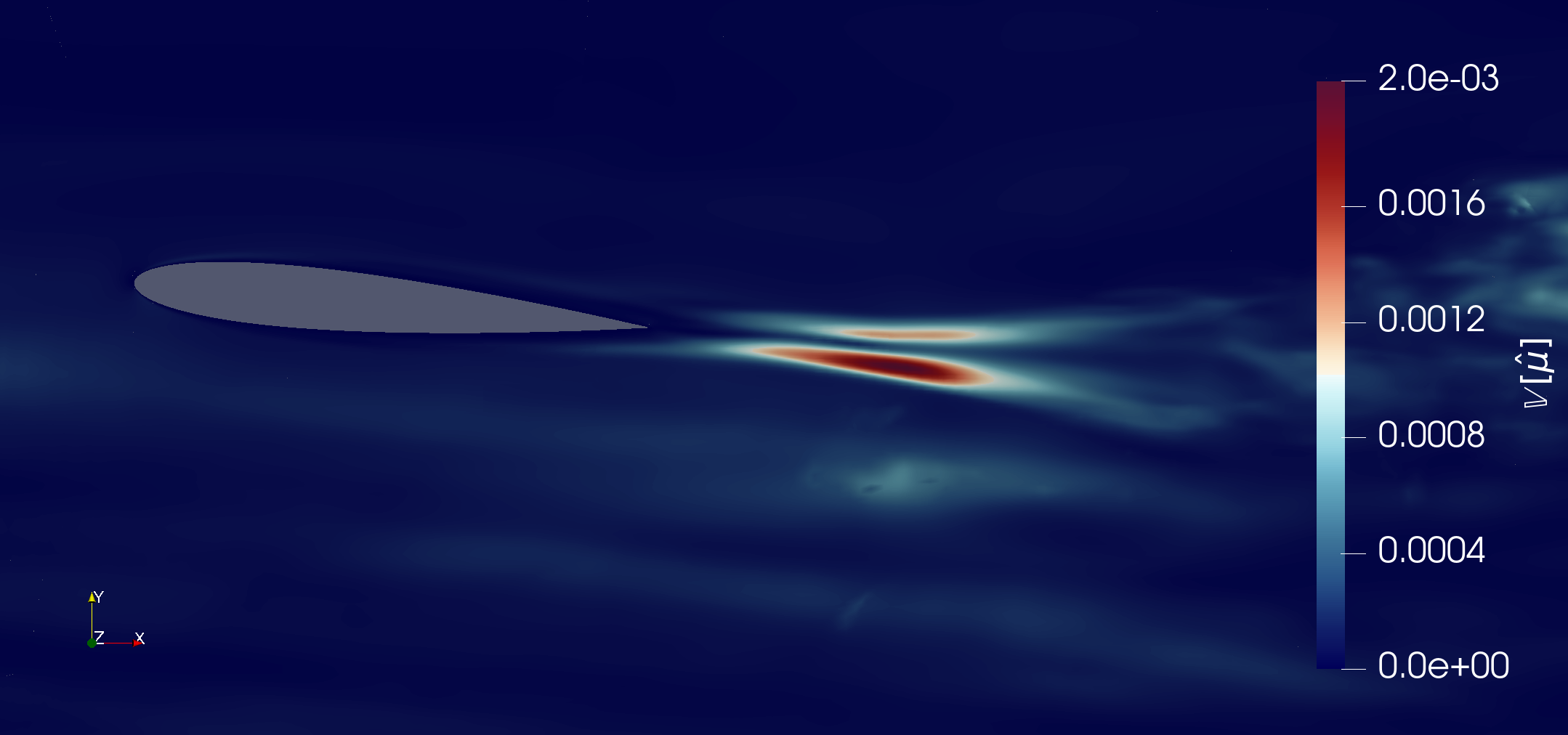}\\
        {\small (c)} & {\small (d)}\\
    \end{tabular}
    \caption{Contours of the SME of the streamwise velocity $u$ of the rotor case at (a) $y=0$ and (b) $z=6$ planes. 
    Corresponding variance of the SMEs obtained from the iMACF method with $M_{\max}=20$ are plotted in (c) and (d), respectively.}\label{fig:rotor_SME_macf}    
\end{figure}

The NACA0012 rotor use case with a 5\textdegree AoA is simulated for a total~$3.6$ time units equivalent to $32400$ time steps with variable size. 
For estimating the uncertainties, the data are interpolated during runtime to equi-distant time steps with a fixed $\Delta t=2\cdot 10^{-3}$ that approximately corresponds to every $18$-th sample of the simulation (\ie~$T_s=18$). 
As a result of this, a total number of $1800$ samples are considered for the iMACF approach and $M_{\max}=20$ is chosen for the UQ estimates. 
The sample-estimated ACF values below $0.4$ are neglected when constructing the ACF model~(\ref{eq:acfModel}). 
The SME of the streamwise velocity component at two planar slices, $y=0$ and $z=6$, together with their uncertainty estimated by the iMACF method are shown in \fig~\ref{fig:rotor_SME_macf}. 
Similar to the NACA4412 wing case, the largest uncertainty can be found in the regions with large turbulence fluctuations and/or long autocorrelations (long-lasting history effects). 
The examples of the latter can be found in the regions of slowly convecting fluctuations or laminar separated flows. 
It should be emphasized that the uncertainty does not necessarily increase with the SME of the velocity, and particularly regions can be identified with SME values close to zero but with high uncertainties, \eg~downstream of the trailing edge of the rotor.

\subsection{Computational performance of the in-situ UQ framework}\label{sec:inSituPerf}
In this section, we discuss the computational performance of the in-situ UQ framework for different mesh sizes and number of MPI ranks considering the NACA4412 wing use case.
The weak and strong scaling tests of the framework are carried out using Nek5000 version~19.0, ParaView Catalyst~5.9 and Python~3.8 on multiple computing nodes that are equipped with~128 physical cores and 256~GB of memory each. 
Based on previous experiences with ParaView Catalyst, we compiled a small edition of the package with only essential functionality that was necessary to execute the UQ framework excluding any rendering features. 
For the simulations, at all points of the 3D mesh, the updating ACF algorithm is applied every~$20$-th time step ($T_s=20$) and with a maximal training lag $m_{\max}=400$ resulting in $20$ different training lags~$\mathbf{m}_{\train}$ to be cached. 
These settings can lead to accurate uncertainty estimates in most practical cases as shown in the previous sections.
The computing time without data I/O and the maximal memory consumption throughout the simulation are measured for different number of MPI ranks, $N_p$, normalized by~$N_{p,ref}=128$. 
The time required to initialize the simulation and to write final results to disk, as well as transition to statistically stationary turbulence were not considered in the tests. 
For each test we ran six simulations, three runs with~$1000$ and three runs with~$2000$ time steps, based on which the computational performance measures with and without UQ were computed as the average time spent per each simulation time step.
In addition, the stand-alone performance of the implementation of \eq~(\ref{eq:varMu}) in Python was investigated offline independent of any CFD code but for the same data matrix sizes provided by the CFD test case. 

\subsubsection{Analytical estimation of the memory consumption}
A lower estimate for the memory consumption of the updating algorithm~\ref{alg:iMACF} can be derived by considering the sizes of the three arrays $M_{i,j}$, $\Gamma_{ij}^m$ and the values of a time series that have to be buffered during runtime. 
For a single flow variable and~$N$ spatial points in the mesh, the memory consumption per rank~$M_r$ can then be estimated by,
\begin{equation}\label{eq:upACF_memory}
    M_r = 3 N (m_{\max}/T_s) S_B \,,
\end{equation}
with~$m_{\max}$ being the maximum lag, $T_s$ the sampling frequency and~$S_B$ the size of the floating point values, usually~$8$ bytes. 
In addition, there is a small memory overhead due to the scalar variables and Python objects that \eq~(\ref{eq:upACF_memory}) does not account for. 
However, for all investigated configurations the overhead is in the order of~$1$ MB and can be neglected compared to the data arrays.
Hence, the maximum lag as well as the sampling interval of the UQ algorithm have a major influence on the memory consumption. 
In simulation cases with large local matrices, these two parameters can be reduced in order to save memory but up to a limit that the uncertainty estimates are still accurate, see \sect~\ref{sec:accuracyMACF}.

\subsubsection{Strong scaling}
\begin{figure}[t]
    \centering
        \begin{tabular}{cc}
        \includegraphics[scale=0.35]{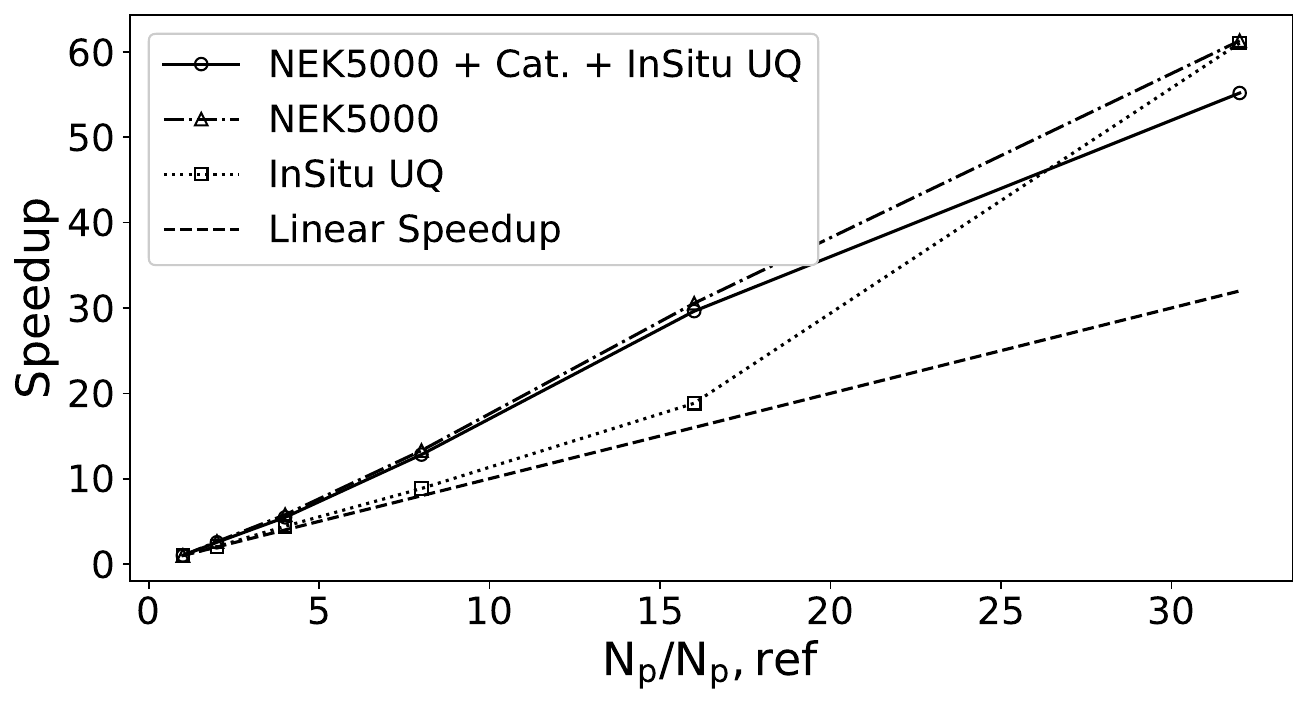} &
        \includegraphics[scale=0.35]{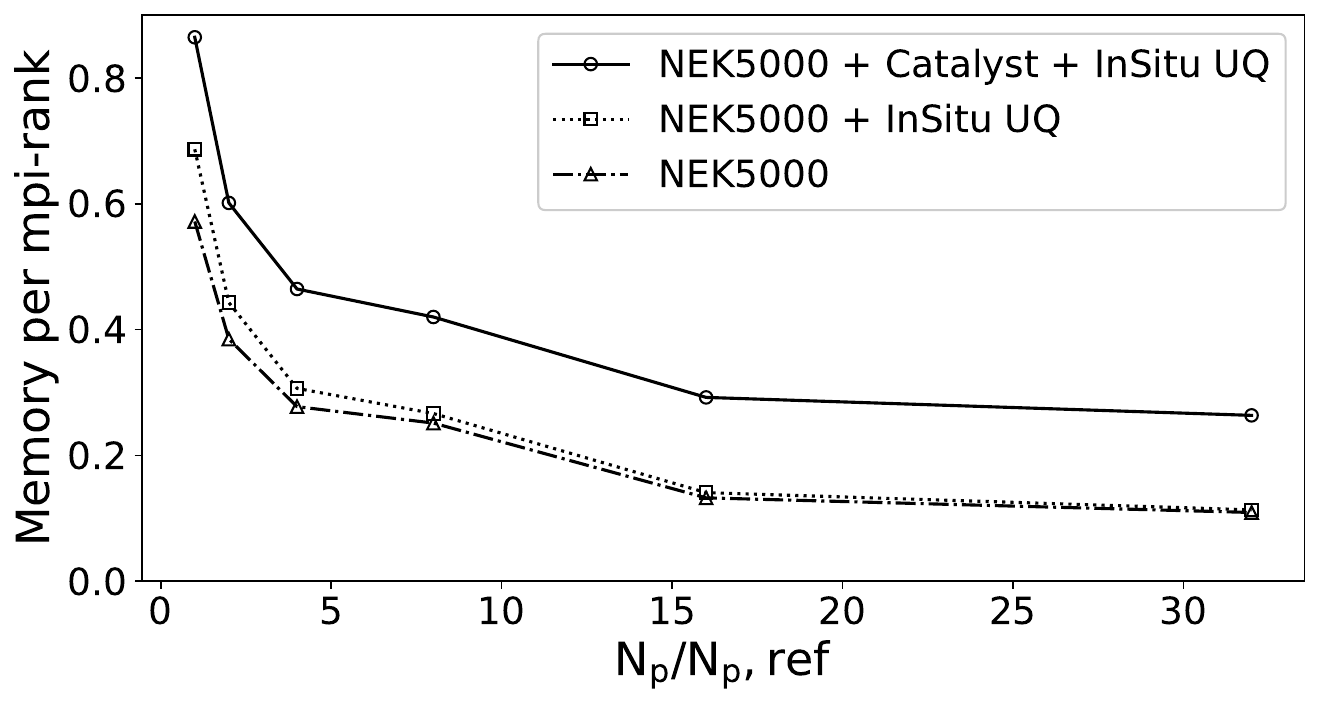}\\
        {\small (a)} & {\small (b)}\\
    \end{tabular}
    \caption{Variation of the (a) speedup and (b) memory consumption (GB) with the number of cores in the strong scaling test. The NACA4412 wing use case is considered with $N_{p,ref}=128$.}
    \label{fig:ss_speedUp_memory}
\end{figure}

For the strong scaling tests of a fixed problem size and increasing number of MPI ranks, see \fig~\ref{fig:ss_speedUp_memory}(a), we observe a super-linear scaling for Nek5000 simulations with and without including the updating UQ framework for all investigated numbers of MPI processors. 
This behavior can be explained by the very small size of the local matrices due to the large number of MPI ranks used. 
In these cases, the matrix sizes can be on the order of the processor's cache size leading to an increase in performance. 
For large local mesh sizes above around~$10000$ points per MPI rank, the entire in-situ framework including Nek5000, the catalyst interface and our UQ code, leads to only a small increase in the overall computing time (less than $5\%$ compared to the UQ-excluded case). 
However, for smaller local meshes the framework does not scale linearly. 
At $N_p/N_{p,ref}=32$ and around $7000$ GLL points per rank, the in-situ UQ framework adds an additional computing time of approximately $11\%$. 
This is most likely due to intermediate steps required for interpolating the data at the GLL points and transferring the mesh from Nek5000 into VTK objects and then to Numpy arrays. 
Therefore, we expect that the performance can be further increased by implementing \eq~(\ref{eq:upACF}) directly into the CFD solver and working on the same data objects as provided there.

An analysis of the memory consumption in the strong scaling tests is shown in \fig~\ref{fig:ss_speedUp_memory}(b). We observe an almost constant overhead of around 160 MB taken up by the Catalyst library with VTK that does not decrease with an increasing number of MPI ranks as the VTK library has to be loaded with each process. 
The updating iMACF algorithm~\ref{alg:iMACF} itself scales well up to large number of MPI ranks.
As compared to the memory consumption of the pure Nek5000 simulations, the iMACF algorithm leads to a $20\%$ increase in memory requirement (144.8 MB per rank) at $N_p/N_{p,ref}=1$ that goes down to~$3 \%$ (4.15 MB per rank) at $N_p/N_{p,ref}=32$. 

\subsubsection{Weak scaling}
In a second study, we evaluate the performance of the in-situ UQ framework for an increasing problem size with a constant number of $42000$ GLL points per rank. 
In order to vary the problem size for these weak scaling tests, we gradually increase the order of the polynomials in each spatial dimension in the spectral elements from~$5$ to~$17$ alongside with increasing the number of MPI ranks from~$54$ to $1458$. 
This increases the total number of GLL points from $2.27$ million to $61.2$ million for changing the polynomials order from~$5$ to $17$.
The computing time is increased by around~$3\%$ after adding the in-situ UQ framework, which is similar to what we observed in the strong scaling tests. 
Both memory consumption per rank and the additional computing time of the framework remain approximately constant over all investigated core counts. This agrees with the expectation, as the updating in-situ algorithm acts only on the local mesh and does not require any intense communication between ranks that would cause a drop in the performance.

\begin{figure}[t]
    \centering
        \begin{tabular}{cc}
        \includegraphics[scale=0.35]{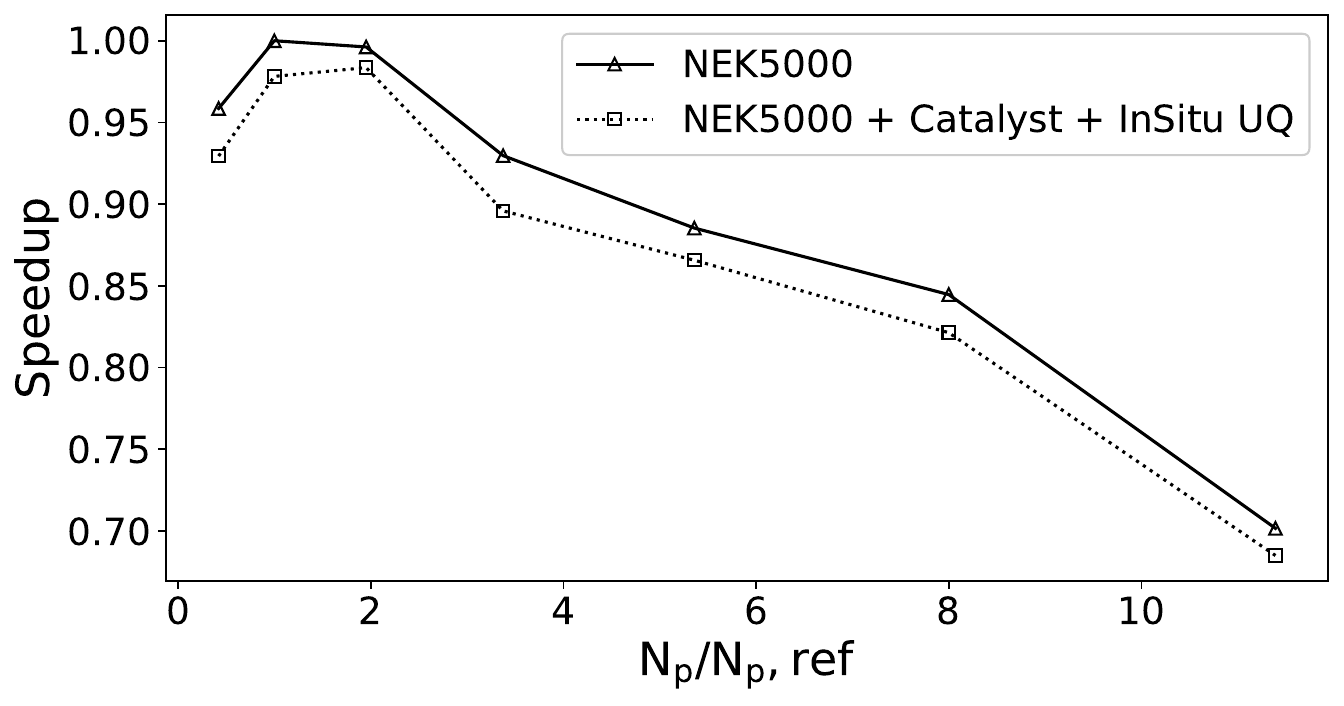} &
        \includegraphics[scale=0.35]{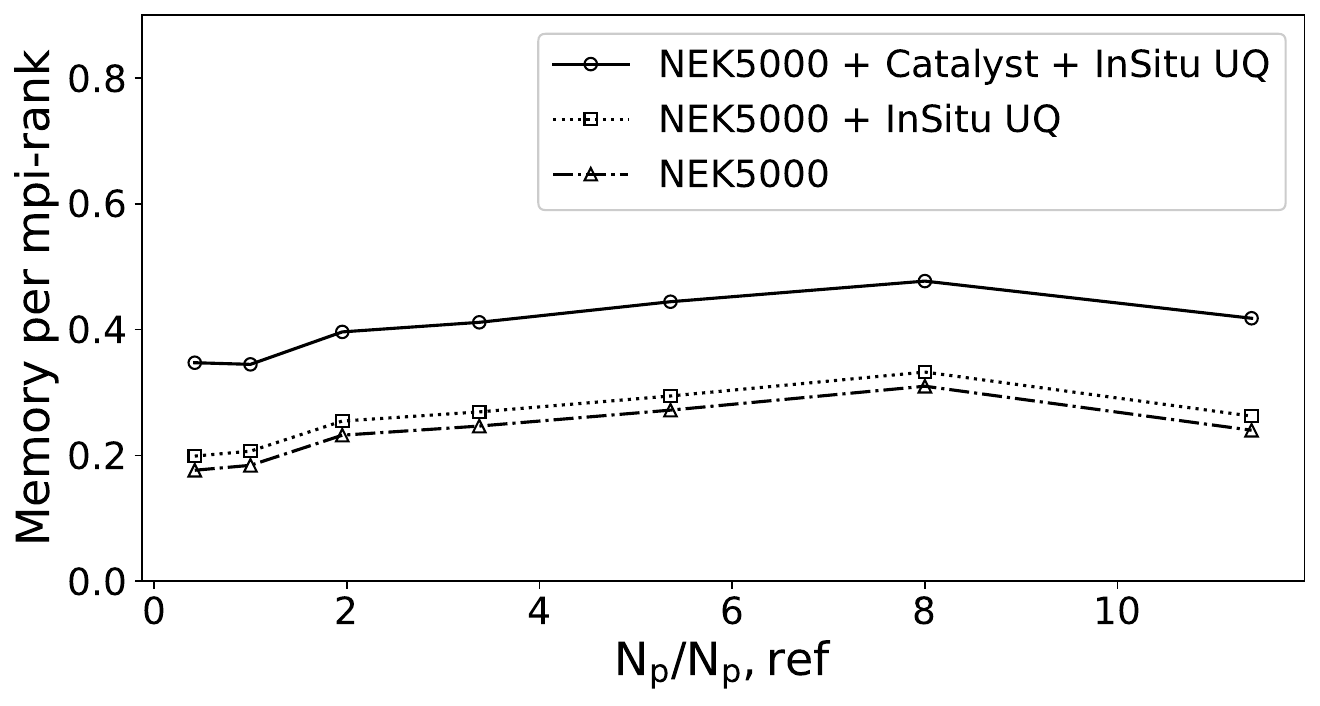}\\
        {\small (a)} & {\small (b)}\\
    \end{tabular}
    \caption{Variation of the (a) speedup and (b) memory consumption (GB) with the number of cores in the weak scaling test. The NACA4412 wing use case is considered with $N_{p,ref}=128$.}
    \label{fig:my_label}
\end{figure}

\section{Summary and Conclusions}\label{sec:summary}
This work introduces a novel framework for in-situ (online/streaming/updating) estimation of time-averaging uncertainties in turbulence statistics. 
This development can be of interest for large-scale simulations of turbulent flows, as it allows for monitoring such uncertainties on-the-fly without any need to store potentially large amounts of time series data on disk. 
A main characteristic of the turbulence time series (signals) considered here is that the samples are autocorrelated up to a generally unknown time lag. 

On the algorithmic side, the present work introduces a streaming formulation for the batch-based methods to estimate time-averaging uncertainties, see \sect~\ref{sec:batchMethods}. 
To remove the natural shortcoming of such approaches, that is the bias in the estimates with respect to the batch size (a user's predefined parameter), we propose a new function, \eq~(\ref{eq:acfModel}), to accurately model the autocorrelation function (ACF) of time series using a limited number of sample-estimated ACFs. 
The resulting modeled ACF is smooth and removes oscillations in the sample ACFs at high time lags.   
Furthermore, we propose an updating formula to evaluate the training sample-estimated ACFs.
These two developments are then integrated into the iMACF method for in-situ estimation of time-averaging uncertainties, see \sect~\ref{sec:iMACF}, where only a minimal memory usage is required. 
A versatile and computationally efficient workflow based on VTK tools is designed and implemented to link an arbitrary CFD solver to the UQ module supplied with the proposed iMACF estimator, see \sect~\ref{sec:workflow}, as well as the updating versions of the non-overlapping batch means (NOBM) and batch-means batch-correlation (BMBC)~\cite{russo:17} methods. 
It is also possible to implement the in-situ \alg~\ref{alg:iMACF} (iMACF method) directly into a CFD solver without any interface.

The accuracy and validity of the MACF and iMACF estimators are thoroughly discussed in \sects~\ref{sec:accuracyMACF} to~\ref{sec:resultsInSitu} using the data from a first-order autoregressive model (ARM) and a fully resolved simulation of a turbulent channel flow. 
According to \fig~\ref{fig:fun1_fun2_toy_channel}, for statistically stationary time series of turbulent channel flow (equivalent to high-order ARMs) the proposed ACF model~(\ref{eq:acfModel}) results in much more accurate modeled ACFs compared to the existing function (\ref{eq:MACF_fun1}) used in literature. 
Moreover, the ACF modeled by function~(\ref{eq:acfModel}) is found to be robust with respect of the choice of the training sample-estimated ACFs, see \fig~\ref{fig:fun2_toy_chan}.
The accuracy of the MACF method is validated by synthetic data, \fig~\ref{fig:pdfMACF_toy}, and compared to the power-law and batch-based methods for turbulent channel flow data, see~\fig~\ref{fig:chan_uncertCompareEstims}.
The detailed investigations represented in \fig~\ref{fig:iMACF_hyperParam_chan} show a negligible influence of the hyperparameters of the iMACF method on the estimated uncertainties as compared to the offline MACF estimator.

In a final step, our new software is successfully applied to a number of more involved use cases including the turbulent flow over a circular cylinder and a wing with structured mesh, as well as a rotor simulated with adaptive mesh refinement.  
In all cases, accurate estimation of uncertainties was obtained by the iMACF method, without any bias that would exist upon using in-situ batch-based methods, see \figs~\ref{fig:naca4412_bmbc} to~\ref{fig:rotor_SME_macf}. 
The detailed analyses in \sect~\ref{sec:inSituPerf}, including the weak and strong scaling tests, prove the efficiency of the framework from both computational time and memory usage aspects.

Although the focus of the present framework is on turbulent flow simulations, the methodology and framework are completely general, thus applicable to statistically stationary time series produced in any application, including laboratory experiments performed using \eg~hotwire anemometry or particle image velocimetry (PIV). 
Future extension of the present framework can be towards including other relevant in-situ data analysis techniques.

\section*{CRediT author statement}
\textbf{SR} and \textbf{CG}: Conceptualization, Methodology, Software, Validation, Formal analysis, Investigation, Writing--Original Draft, Visualization.
\textbf{AP}: Investigation, Writing--Original Draft. 
\textbf{JG}: Writing--Review \& Editing, Supervision, Funding acquisition. 
\textbf{PS}: Conceptualization, Writing--Review \& Editing, Funding acquisition.

\section*{Acknowledgments}
This work has been supported by the EXCELLERAT project which has received funding from the European Union's Horizon 2020 research and innovation programme under grant agreement No 823691.
This work has received funding from the European High Performance Computing Joint Undertaking (JU) and Germany, Italy, Slovenia, Spain, Sweden, and France under grant agreement No 101092621.
The authors would like to thank Dr. Marco Atzori from The Polytechnic University of Milan, Italy, for providing the Nek5000 case for the wing simulation, and  Dr. Antonio Memmolo from CINECA, Italy, for sharing the data of the "toy" rotor simulation.
The computation of the rotor case was enabled by resources provided by the Swedish National Infrastructure for Computing (SNIC) at the PDC Center for High Performance Computing, KTH Royal Institute of Technology, partially funded by the Swedish Research Council through grant agreement no. 2018-05973.
The simulations of the NACA4412 wing case were carried out on Hawk at the High Performance Computing Center Stuttgart (HLRS). Funding for Hawk was provided by Baden-Württemberg Ministry for Science, Research, and the Arts and the German Federal Ministry of Education and Research through the Gauss Centre for Supercomputing (GCS).

\appendix

\section{The Power-law Method to Model an ACF}\label{app:ARM_modeledACF}
A smooth model for ACF, to be plugged into \eq~(\ref{eq:varMu}), can be obtained by first fitting an autoregressive model (ARM) of order~$p$ to samples $\{x'_i\}_{i=1}^n$ where $x'_i=x_i-\hmu_x$~\cite{wei:06}:
\begin{equation}\label{eq:ARM}
    x'_i=\sum_{j=1}^p \alpha_j x'_{i-j} + \epsilon_i\,.
\end{equation}
The noise term is Gaussian, $\epsilon_i\sim \mathcal{N}(0,\sigma_d^2)$ where the standard deviation~$\sigma_d$ and the coefficients~$\{ \alpha_j \}_{j=1}^p$ can be obtained by various methods, see~\cite{wei:06}. 
The starting point for deriving a smooth model for ACF is the Yule-Walker equation, see~\cite{wei:06}:
\begin{equation}\label{eq:yw}
    \rho_k = \sum_{j=1}^p \alpha_j \rho_{k-j}\,, \quad 
    k=1,2,\cdots,p \,,
\end{equation} 
where $\rho_k$ is the autocorrelation of~$x$ at lag~$k$.
Plugging the ansatz $\rho_k=\lambda^k$ in this equation leads to an algebraic characteristic equation with roots $\{\lambda_k\}_{k=1}^p$ by which the ACF at lag $k$ can be expanded as, 
\begin{equation}\label{eq:ARM_ACF}
    \rho_k= \sum_{i=1}^p c_i \lambda_i^k \,.
\end{equation}
To compute the coefficients~$\{c_i\}_{i=1}^p$ from a linear system of equations, a set of~$K_{ACF}$ sample-estimated ACF at lower lags (to avoid the issue of wiggles at high lags) are considered in \eq~(\ref{eq:ARM_ACF}). 
This results in a smooth model for the ACF at any lag $k\geq 0$.
The utilization of modeled ACF in \eq~(\ref{eq:varMu}) is referred to as the ARM-based or power law uncertainty estimation method.
The hyperparameters of this method are the ARM order~$p$ and the set of training sample-estimated ACFs with size $K_{ACF}$. 

\section{Reduction of \eq~(\ref{eq:upACF}) to \eq~(\ref{eq:upS})}\label{app:acf2var}
By setting $m=0$, \ie~lag zero in \eq~(\ref{eq:upACF}), we can recover \eq~(\ref{eq:upS}). The main steps of the derivation are given below. 
\begin{eqnarray}
\Gamma_{i,j}^0 &=& 
\Gamma_{i,j-1}^0 
- \Delta M_{i,j}\sum_{k=i}^{j-1} 2x_k 
+ (j-i+1)M_{i,j}^2
- (j-i)M_{i,j-1}^2
+ x_j^2 -2x_jM_{i,j} \nonumber\\
&=&\Gamma_{i,j-1}^0 
- 2(j-i)\Delta M_{i,j}M_{i,j-1}
+ (j-i)M_{i,j}^2
+ (M_{i,j}-x_j)^2
- (j-i)M_{i,j-1}^2 \nonumber\\
&=&\Gamma_{i,j-1}^0 
- 2(j-i)\Delta M_{i,j}M_{i,j-1}
+ (j-i)(M_{i,j-1}+\Delta M_{i,j})^2
+ (j-i)^2\Delta M_{i,j}^2
- (j-i)M_{i,j-1}^2 \nonumber \\
&=&\Gamma_{i,j-1}^0
+ (j-i)(j-i+1)\Delta M_{i,j}^2 \,,
\end{eqnarray}
where $\Gamma_{i,j-1}^0=S_{i,j-1}$.

\section{Impact of the Batch Size on the Batch-based Uncertainty Estimators}\label{app:batchEffect}
\begin{figure}[t]
    \centering
    \begin{tabular}{c}
         \includegraphics[scale=0.5]{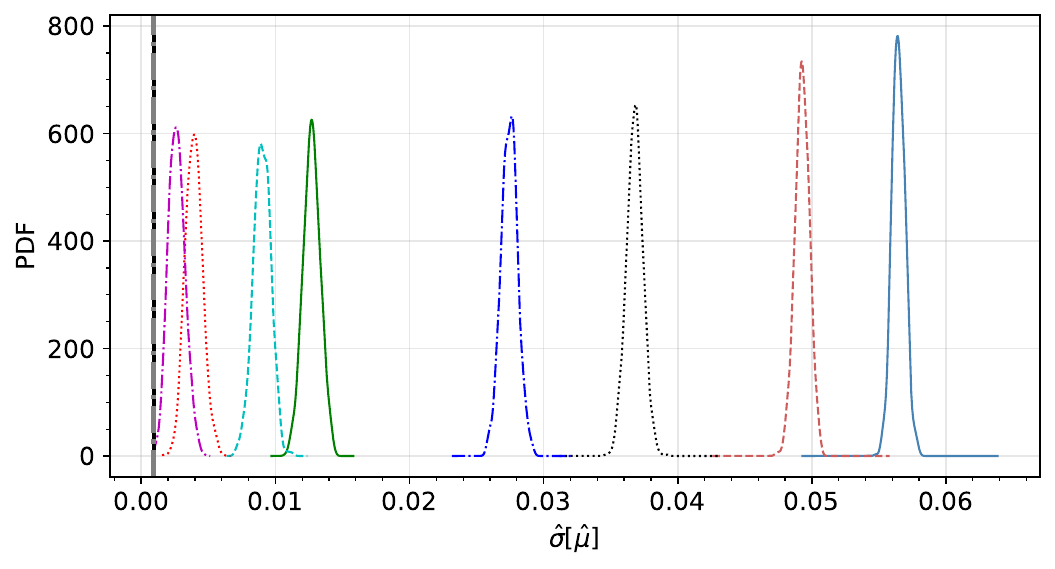}\\
         \includegraphics[scale=0.5]{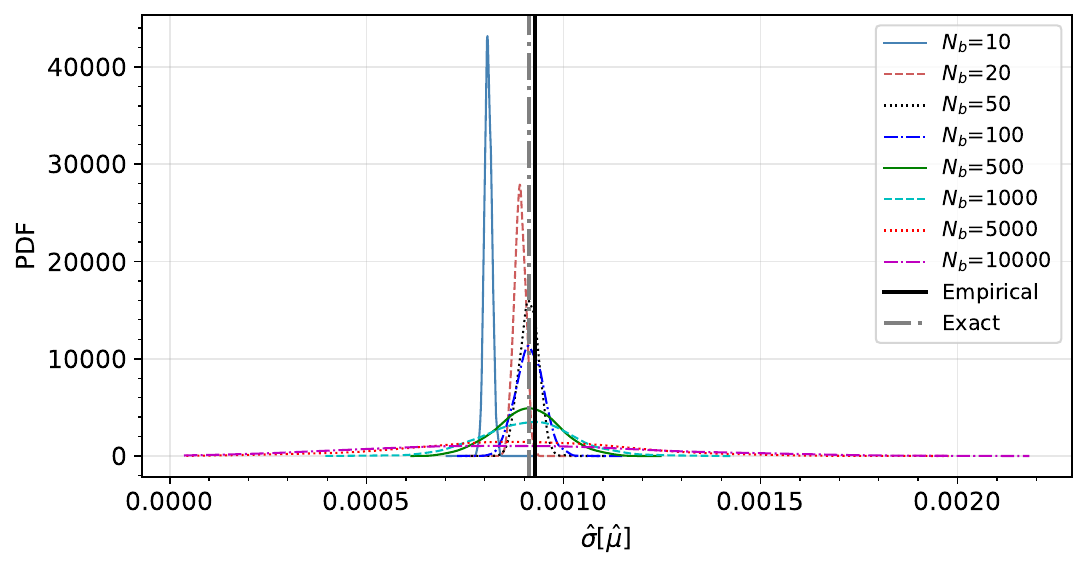}\\
    \end{tabular}
    \caption{Impact of the batch size on the estimated $\hsig(\hmu)$ for the synthetic samples generated by the ARM(1) defined in \eq~(\ref{eq:ARM1}) for (top) NOBM and (bottom) BMBC methods. The PDFs are obtained by $1000$ repetitions of independent simulations with the number of samples $n$ in each simulation being equal to $10^5$.}
    \label{fig:batchSize}
\end{figure}

The batch-based methods are the most frequently used approaches for estimating time-averaging uncertainties in autocorrelated turbulence time series. 
However, the batch size introduces a bias in such estimated uncertainties~\cite{xavier:21}. 
This is shown here by adopting the procedure described in \sect~\ref{sec:validMACF} with the samples generated from the first-order ARM defined in \eq~(\ref{eq:ARM1}). 
\fig~\ref{fig:batchSize} represents the PDF of the SME's uncertainties obtained by the non-overlapping Batch Means (NOBM) and the Batch-Means Batch-Correlation (BMBC) methods proposed in \rf~\cite{russo:17}.
In both plots, the empirical and exact uncertainties in the SME are also shown. 
The bias introduced by the batch size is clearly more evident for the NOBM method, but it is interesting that even for ARM(1), the BMBC method is still affected. 
Another important observation is that for the NOBM method, the confidence in the estimated~$\hsig(\hmu)$ is not changing significantly with the variation of the batch size, but for the BMBC method, the confidence in the estimations is reduced as the batch size increases. 
The latter is due to the reduction of the number of batches as the batch size increases while the total number of available samples is fixed. 
This issue is the main barrier to the use of batch-based methods in an in-situ way, noting that no valid estimates can be made for the optimal batch size for the time series prior to the simulations.
Furthermore, any preset batch size may not be applicable for all quantities of interest and the spatial points in a region of interest, noting the dependence of the time series and associated ACF to these factors. 

\bibliographystyle{abbrvnat}  
\bibliography{bib_InSituUQ}

\end{document}